\documentclass[twocolumn]{aastex7}

\usepackage{tipa}

\usepackage{booktabs}
\usepackage{siunitx}
\newcommand{\h}{$^{\mathrm{h}}$}
\newcommand{\m}{$^{\mathrm{m}}$}
\newcommand{\seco}{$^{\mathrm{s}}$}

\newcommand{\rh}{$r_{\text{h}}$}
\newcommand{\rhm}{r_{\text{h}}}

\newcommand{\Msun}{$\textup{M}_\odot$}
\newcommand{\Lsun}{$\textup{L}_\odot$}
\newcommand{\sqdeg}{deg$^2$}
\newcommand{\kms}{km\,s$^{-1}$}
\newcommand{\met}{[Fe/H]}

\newcommand{\uii}{UNIONS 2}
\newcommand{\uiii}{UNIONS 3}
\newcommand{\boovi}{Bo\"otes VI}
\newcommand{\draiii}{Draco III}

\begin{document}



\title{State of the UNIONS: \\ Four Milky Way companions discovered in a systematic search for northern Local Group satellites}

\author[orcid=0000-0002-6946-8280,sname='North America']{Simon E.\,T. Smith}
\affiliation{Department of Physics and Astronomy, University of Victoria, Victoria, BC, V8P 1A1, Canada}
\email[show]{simonsmith@uvic.ca}  

\author[orcid=0000-0003-4666-6564]{Alan W. McConnachie}
\affiliation{NRC Herzberg Astronomy and Astrophysics, 5071 West Saanich Road, Victoria, BC, V9E 2E7, Canada}
\affiliation{Department of Physics and Astronomy, University of Victoria, Victoria, BC, V8P 1A1, Canada}
\email[]{}

\author[orcid=0000-0001-5486-2747]{Thomas de Boer}
\affiliation{Institute for Astronomy, University of Hawaii, 2680 Woodlawn Drive, Honolulu HI 96822}
\email[]{}  

\author[orcid=0000-0001-6965-7789]{Ken C. Chambers}
\affiliation{Institute for Astronomy, University of Hawaii, 2680 Woodlawn Drive, Honolulu HI 96822}
\email[]{}

\author[orcid=0000-0001-8221-8406]{Stephen Gwyn}
\affiliation{NRC Herzberg Astronomy and Astrophysics, 5071 West Saanich Road, Victoria, BC, V9E 2E7, Canada}
\email[]{}

\author[orcid=0000-0002-9814-3338]{Hendrik Hildebrandt}
\affiliation{Ruhr University Bochum, Faculty of Physics and Astronomy, Astronomical Institute (AIRUB), German Centre for Cosmological Lensing, 44780, Bochum, Germany}
\email[]{}

\author[orcid=0000-0002-4350-7632]{Jaclyn Jensen}
\affiliation{Department of Physics \& Astronomy, Dartmouth College, 6127 Wilder Laboratory, Hanover, NH 03755, USA}
\affiliation{Department of Physics and Astronomy, University of Victoria, Victoria, BC, V8P 1A1, Canada}
\email[]{}

\author[orcid=0000-0002-7965-2815]{Eugene Magnier}
\affiliation{Institute for Astronomy, University of Hawaii, 2680 Woodlawn Drive, Honolulu HI 96822}
\email[]{}

\author[orcid=0000-0002-1349-202X]{Nicolas F. Martin}
\affiliation{Universit\'e de Strasbourg, CNRS, Observatoire Astronomique de Strasbourg, UMR 7550, F-67000 Strasbourg, France}
\affiliation{Max-Planck-Institut f\"ur Astronomie, K\"onigstuhl 17, D-69117, Heidelberg, Germany}
\email[]{}

\author[orcid=0000-0002-7507-5985]{Akshara Viswanathan}
\affiliation{Department of Physics and Astronomy, University of Victoria, Victoria, BC, V8P 1A1, Canada}
\email[]{}

\author[orcid=0009-0000-6226-9345]{Shiyang Zhang}
\affiliation{Ruhr University Bochum, Faculty of Physics and Astronomy, Astronomical Institute (AIRUB), German Centre for Cosmological Lensing, 44780, Bochum, Germany}
\email[]{}

\begin{abstract}

We present four newly discovered Milky Way satellites from a systematic search through the {\it gri} photometry from the Ultraviolet Near-Infrared Optical Northern Survey (UNIONS). Two of these satellites are likely ultra-faint dwarf galaxies that we have named Bo\"otes VI and Draco III. 
Both galaxy candidates are $\gtrsim 100$\,kpc from the Galactic center and have absolute $V$-band magnitudes ($M_V$) of $-2.7^{+0.6}_{-0.5}$ and $-4.0\pm0.6$, respectively. Bo\"otes VI and Draco III have physical half-light radii ($r_{\rm h}$) of $\sim 50$\,pc, or about twice as large as the most extended Milky Way globular clusters with similar luminosities. They are therefore candidate galaxies, though chemodynamical analyses will be needed to secure these classifications. 
We name the two other systems UNIONS 2 and UNIONS 3. They are extremely faint ($M_V = +0.4^{+0.6}_{-0.9}$ and $-0.7^{+0.6}_{-0.8}$) and compact ($r_{\rm h} = 5\,\pm\,1$ and $10^{+3}_{-2}$\,pc), and zero giant stars are identified in either of their present-day stellar populations.
Morphological measurements alone are not sufficient to assess the presence of dark matter in either system, so do not provide definitive classifications. 
We discuss the challenges in either detecting or refuting the presence of dark matter in UNIONS 2 and UNIONS 3, two systems that are archetypal of the faintest satellites that will be discovered in increasing numbers in upcoming wide-field surveys.


\end{abstract}


\section{Introduction} 
\label{sec:intro}

In the Milky Way halo, the faintest confirmed satellite galaxies known have total luminosities less than 300\,\Lsun\ \citep[Tucana V, Segue 1, Triangulum II;][]{Simon20, Munoz18, Richstein24}. These are the most extreme examples from a class of galaxies referred to as ultra-faint dwarfs (UFDs), defined as galaxies with $M_V > -7.7$ \citep{Bullock17, Simon19}. They are ancient (age, $\tau \gtrsim 10$\,Gyr), metal-poor ($\langle{\rm [Fe/H]}\rangle \lesssim -2$), and have large implied dynamical mass-to-light ratios (M$_{\rm dyn}$/L $\gtrsim 10^3$\,\Msun/\Lsun).

In the context of $\Lambda$ Cold Dark Matter ($\Lambda$CDM) cosmology, UFDs are highly dark matter dominated and situated as the lowest mass fossils of galaxy formation in the early universe \citep{Bovill11, Brown12, Brown14, Frebel14}. Individual UFDs have become premier laboratories for investigating baryonic processes such as self-enrichment \citep{Kirby13, Luna25}, chemical evolution \citep[e.g.][]{Frebel10, Ji16, Ji19, Roederer16}, the local and environmental impacts of reionization \citep{Bullock00, Ricotti05, Bovill09}, and low mass galaxy mergers \citep{Deason14, Chiti21, Waller23}, as well as open questions in dark matter theory, like the shape and density of small dark matter halos \citep[e.g.][]{SanchezAlmeida25}, and their modification in a halo environment \citep{Errani21, Errani22, Errani24-micro}. On a population level, the number and distribution of UFDs has spurred progress in the understanding of the galactic luminosity function \citep{Simon07, Koposov08, Tollerud08, Walsh09, Newton18, Drlica20}, the halo occupation function \citep{Jethwa18, Nadler20, Santos25} and the viability of alternate dark matter models \citep{Nadler19, Newton21, Deason22, McDaniel24}, as well as tidal stripping \citep{Errani21, Jensen24, Riley25} and the accretion history of the Milky Way \citep{Deason16, Santos25, Shipp25}.

The challenge of uncovering galaxies at ever fainter limits, and thereby extending galactic evolution models to lower masses and smaller length-scales, has been enabled by the era of digital sky surveys and wide-field photometry performed over the extragalactic sky. Only nine ``classical'' dwarf galaxies ($M_V < -7.7$) were known prior to the turn of the millennium. The Sloan Digital Sky Survey \citep[SDSS,][]{York00, Abazajian09} saw $\sim$15 Milky Way satellite dwarf galaxies discovered \citep[e.g.][]{Willman05-uma, Willman05-w1, Belokurov06-boo, Belokurov07-quin, Belokurov09, Belokurov10, Zucker06-uma, Zucker06-cvn1}, revealing the first UFDs and more than doubling the known population of Milky Way dwarf galaxies. The Dark Energy Survey \citep[DES,][]{Abbot18} and the 3$\pi$ survey with the Panoramic Survey Telescope and Rapid Response System \citep[Pan-STARRS,][]{Chambers16} enabled the discovery of another $\sim$25 UFDs \citep{Bechtol15, Drlica15, Koposov15, Laevens15-one, Laevens15-three}. Most recently, the DECam Local Volume Exploration survey \citep[DELVE,][]{Drlica21, Drlica22} has been particularly productive hunting grounds \citep{Mau20, Cerny21, Cerny23-peg4, Cerny23-d6, Cerny25, Tan25-leo, Tan25-three} while additional discoveries \citep{Homma16, Homma18, Homma19, Homma2024, Gatto2024, Smith23} in the Hyper-Suprime-Cam Subaru Strategic Program \citep[HSC-SSP,][]{Aihara18-overview, Aihara18-Y1} and the Ultraviolet Near-Infrared Northern Survey \citep[UNIONS,][]{Gwyn25} have brought the total number of confirmed or suspected Milky Way satellite UFDs to 57 \citep{Pace25, Cerny26-chem}. 

Looking beyond the Milky Way, our largest neighbor Andromeda (M31) hosts a larger population of dwarf galaxies per luminosity bin, with 43 galaxies known brighter than $M_V = -5$, 15 of which are UFDs. The Pan-Andromeda Archaeological Survey \citep[PAndAS;][]{McConnachie2009} discovered many of M31's satellites within 150\,kpc while the outer region of the halo population has been added to by SDSS, Pan-STARRS, UNIONS, and the DESI Legacy Imaging surveys \citep[][]{Dey2019}. In the rest of the Local Group, the galactic region surrounding the Milky Way and M31, there are roughly 65 known galaxies out to 3\,Mpc. Only a handful are UFDs ($\sim 5$), reflecting the difficulty of detecting the the lowest mass, least evolved galaxies beyond the halo environment.

Returning to the Milky Way halo, the dark matter-dominated dwarf galaxies are contrasted by nearly 200 self-gravitating star clusters (SCs), the majority of which ($\sim$150) are the classical globular clusters \citep[GCs,][2010 edition]{Harris96}. Most GCs can be easily distinguished from dwarf galaxies via imaging: in the magnitude range of typical GCs ($-10 < M_V < -5$), they are far more compact (half-light radii, \rh\ $\lesssim 10$\,pc in scale radius) than dwarfs (\rh\ $\gtrsim 100$\,pc). Stellar spectroscopy has also revealed key differences: line-of-sight velocities imply that GCs have $1 <$ M$_{\rm dyn}$/L $< 3$\,\Msun/\Lsun\ \citep[e.g.][]{Baumgardt20} and chemical analyses show little spread in iron abundance \citep[e.g.][]{Leaman12}, implying short star formation histories and negligible self-enrichment \citep{Baumgardt08, Willman12}. This difference is thought to reflect the fundamental distinction between the populations, namely that dwarf galaxies form at the centers of their own dark matter halos, whereas GCs do not \citep[see][and citations therein]{Kruijssen26}.

The faint end of the confirmed UFD luminosity function tends to smaller physical sizes and merges with the faint tail of the GCs luminosity function, challenging the use of morphological features alone to classify newly discovered satellites as dark matter-free SCs or dark matter-dominated UFDs \citep[e.g.][]{Willman05-w1, Misgeld11}. In the region of size-luminosity space roughly defined by $M_V > -3$ and $r_{\rm h} < 20$\,pc, there are $\sim$\,30 ultra-faint compact satellites \citep[UFCSs,][]{Cerny26-chem}\footnote{Highlighted by \citep{Pace25}, the literature shows a lack of agreement surrounding these systems, as they have been called many names including ultra-faint star clusters \citep{Kim16, Cerny23-six} ultra-faint objects \citep{Contenta17}, faint halo clusters \citep{Kim15-outer}, faint ambiguous satellites \citep{Smith24}, hyper-faint compact stellar systems \citep{Pace25}, and glorfs \citep{Errani24-uma}} of the Milky Way whose natures remain ambiguous. This is in part due to a historical quirk; the first of these systems to be spectroscopically explored (Segue 3 and Muñoz 1) showed no clear positive evidence for dark matter, so as additional compact systems were discovered, they were typically assumed to be SCs and de-prioritized for spectroscopic follow-up \citep{Fadely11, Munoz12}.

The functional challenge when studying UFCSs is in obtaining sufficiently large spectroscopic samples of member stars in extra-galactic systems with few-to-no giant stars in their present day stellar populations \citep[e.g.][]{Kim15-hero, Kim16, Homma19}. The lack of targets severely hinders the ability to measure a reliable dynamical mass or metallicity spread that may indicate the presence of a dark matter halo \citep{Cerny26-chem}. 
It should be noted that systems that fall under the UFCS heading are not necessarily a physical class distinct from UFDs or GCs. Rather the term UFCS is used to hold a place for satellites whose natures are ambiguous {\it given present data}, as presumably each system will prove to contain dark matter or not when the appropriate observations are obtained. 
Only recently, due to renewed interest and effort, have some of these previously known ambiguous systems been shown to exhibit the chemodynamical signatures that may place them as the faintest, most compact galaxies in existence \citep{Errani24-uma, Simon24, Smith24, Cerny26-chem}.

In this manuscript, we present the results of a search in the newest data from UNIONS, and the discovery of four Milky Way satellites. We present two candidate UFDs that are more than 2 times more physically extended than the most extended GCs, named \boovi\ and \draiii, and two extremely faint, compact satellites whose natures remain ambiguous at this time, named \uii\ and \uiii. Sec. \ref{sec:data} introduces the data used in this search while Sec. \ref{sec:search} outlines our isochrone-filter search methodology and survey-level results. Sec. \ref{sec:methods} describes every measurement applied to each newly discovered satellite, the results of which are presented in Sec. \ref{sec:results}. We discuss the classification of each system as well as the prospects for determining the nature of the two ambiguous systems in Sec. \ref{sec:discussion}. We conclude the manuscript in Sec. \ref{sec:conclusions}.

\section{Data}
\label{sec:data}

\subsection{UNIONS}
\label{subsec:unions}

UNIONS
is a wide-field digital survey conducted in five broad photometric bands ($ugriz$) across the northern hemisphere \citep{Gwyn25}. 
UNIONS is a ``collaboration of collaborations'', wherein the Canada-France-Hawai'i Telescope (CFHT), the Pan-STARRS telescopes, and the Subaru Observatory each contribute imaging to the consortium: the {\it Canada-France Imaging Survey} \citep[CFIS;][]{Ibata17} is obtaining $u$- and $r$-band imaging with MegaCam on CFHT; Pan-STARRS is obtaining $i$- and $z$-band imaging; the HSC at Subaru is obtaining $z$-band imaging through the {\it Wide Imaging with Subaru HSC of the Euclid Sky} (WISHES) and $g$-band imaging through the {\it Waterloo-Hawai'i Institute for Astronomy g-band Survey} (WHIGS).
UNIONS is a vital contributor to the ground-based component of the \textit{Euclid} space mission \citep{Euclid25-overview}. The primary footprint of the survey is roughly 4600\,\sqdeg\ of northern sky ($\delta \geq +30$\textdegree) away from the Galactic plane ($|b| \geq 25$\textdegree) in the North Galactic Cap (NGC). Additional data collection is underway to extend the NGC coverage by $\sim1400$\,\sqdeg\ to fill a strip of sky between $+30 \geq \delta \geq +15$\textdegree\ \citep[See Sec. 2.3.1 in][]{Gwyn25} to complement the coverage of the Rubin Observatory's Legacy Survey of Space and Time \citep[LSST,][]{Ivezic19}.
The secondary field is $\sim 300$\,\sqdeg\ in the South Galactic Cap (SGC) towards the Andromeda galaxy meeting the $\delta \geq +30$\textdegree\ and the same $b$ constraints as the NGC. 
The final coverage of UNIONS will consist of nearly $6500$\,\sqdeg\ of extragalactic sky, surveyed in five photometric bands.

\subsection{Data used in this work}
\label{sec:thisdata}

Each constituent survey in UNIONS employs different observing strategies and photometric techniques that manifest as variable sky coverage and depth across the nominal UNIONS footprint at the time being. This work uses the internal catalogues available to the collaboration as of June 2025. Here, we summarize these catalogues, and we refer the reader to \citet{Gwyn25} for a full discussion of the image processing, photometric catalogue pipelines, and planned future public data releases.

\begin{figure*}
    \centering
    \includegraphics[width=\linewidth]{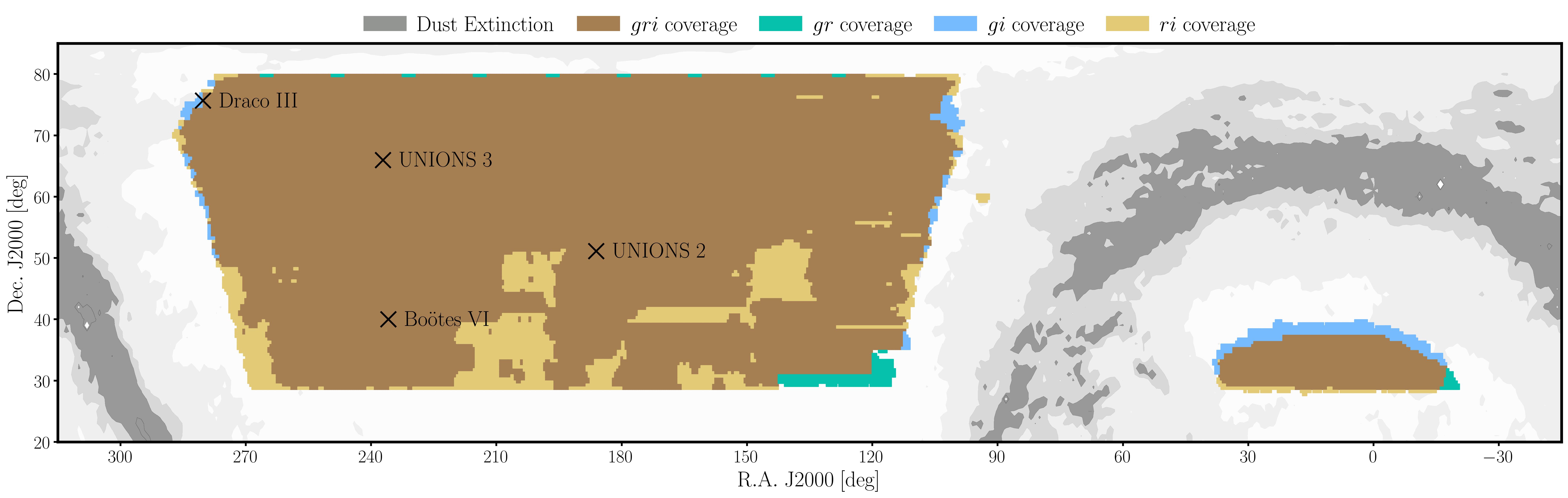}
    \caption{Sky coverage of the UNIONS data used in this paper. Due to the nature of UNIONS as a ``collaboration of collaborations", the overlap between the $g$, $r$, and $i$ catalogue data varies across the final footprint in the current set of internal data releases. The overlap of each photometric pair is shown and color-coded, along with the total $gri$ combined coverage. The four newly discovered Milky Way satellites presented in this work are overlaid as black X's. The grey contours show galactic extinction at levels of $E(B-V) = [0,\,0.1,\,0.5,\,1,\,5,\,10]$\,mag, taken from \citet{Schlegel98}.}
    \label{fig:unions-cov}
\end{figure*}

We have carried out our search for Local Group satellites in paired combinations of $gri$ photometric catalogue data. Each pair of photometric catalogues has the following total areal overlap: $gr$ covers 4361.75\,\sqdeg, $gi$ covers 4378.5\,\sqdeg, and $ri$ covers 5149.25\,\sqdeg. We display the spatial overlap in Fig. \ref{fig:unions-cov}. We have also selected the source magnitudes most appropriate for stellar photometry from each catalogue, namely $g$-band point spread function (PSF) magnitudes from WHIGS, $r$-band 1-D radially averaged PSF magnitudes from CFIS, and $i$-band PSF magnitudes from Pan-STARRS, all of which are measured on the stacks. For the overdensity search in Sec. \ref{sec:methods}, we de-extinct the raw photometry using the maps from \citet{Schlegel98} and the broadband coefficients from \citet{Schlafly11}. When examining individual satellites, we then re-calibrate \citet{Schlegel98} following the 14\% adjustment suggested by \citet{Schlafly11}.

For the $gr$ and $ri$ combinations, we separate stars and galaxies morphologically using the $r$-band due to its excellent image quality. The median full-width-at-half-maximum of the seeing disk is 0.7$\arcsec$. Stars are well separated from galaxies down to $r \sim 23.2$\,mag, below which there is considerable contamination from background galaxies (Thomas et al. in prep.). For the $gi$ combination, we rely on the $g$-band for morphological classification, which is processed using the LSST Science Pipeline \citep{Bosch18, Bosch19} and we select point sources by taking $\texttt{base\_ClassificationExtendedness\_value} = 0$  \citep[equivalent to \texttt{classification\_extendedness} flag, see Sec. 4.9.10 in][]{Bosch18}.

We compare a 10\,\sqdeg\ patch of the CFIS $r$-band data with deeper $r$-band data from HSC-SSP to estimate the depth at which incompleteness becomes severe for stars selected from both surveys. We find that completeness per magnitude bin is very high $>0.95$ for bright magnitudes. As a function of $\sigma$ point source depth, we find that completeness in $r$ is 0.94 at 15$\sigma$, falling to 0.78 at 10$\sigma$, and dropping sharply to 0.13 at 5$\sigma$. We use this as a rough guide for all bands. 
For the dwarf galaxy search (Sec. \ref{subsec:isochrone-filter}), we prioritize taking advantage of the excellent depth of UNIONS and therefore use the 10$\sigma$ depth as a faint-end magnitude cut. Across the entire UNIONS footprint, these we estimate 10$\sigma$ occurs at [$g$, $r$, $i$] $\simeq$ 24.6, 24.9, and 23.4\,mag. When analyzing individual candidate satellites, we prioritize completeness by restricting our photometry to 15$\sigma$ depths which are measured for each satellite and given in Table \ref{tab:final-properties}.

\section{Satellite Search}
\label{sec:search}

\subsection{Isochrone-Filter Search}
\label{subsec:isochrone-filter}

We search for dwarf galaxies using the isochrone-filter method which has been successfully implemented in many forms to find satellites of the Milky Way \citep[e.g.][]{Koposov08, Walsh09, Bechtol15, Koposov15, Laevens15-three}, and M31 \citep[e.g.][]{McConnachie08, Richardson11, Slater11, Martin13-bayes, Martin13-2new}. This methodology has been previously applied to earlier internal data releases from UNIONS, leading to the discoveries of the candidate Milky Way dwarf galaxy Bo\"otes V \citep[$M_V = -3.2$,][]{Smith23}, the faintest known satellite of the Milky Way Ursa Major III/UNIONS 1 \citep[$M_V = +2.2$,][]{Smith24}, and a faint, distant member of the M31 sub-group Pegasus VII \citep[$M_V = -5.7$,][]{Smith25}. The specific implementation is given in \citet{Smith23}, but we briefly outline it here.

Using each pair of photometric bands, we select stars whose colors are within 0.1\,mag of an interpolated 12\,Gyr, [Fe/H] = $-2$\,dex PARSEC isochrone \citep{Bressan12} placed at a test distance $d$. We also add the typical photometric uncertainties of each band to the 0.1\,mag width in quadrature, such that at faint magnitudes, the width of the selection region grows in size with the characteristic increase in photometric uncertainty. For each iteration, we shift the isochrone to increasing heliocentric distances that range from 10\,kpc to 1000\,kpc spaced roughly logarithmically. We also apply a magnitude cut-off at the global 10$\sigma$ point-source depth of the $g$- and $r$-bands. Selected stars are binned on a 1\arcmin\,$\times$\,1\arcmin\ grid in tangent plane-projected coordinates where we convolve with a small 1.5\arcmin\ Gaussian kernel to produce a sky map of stellar density, as well as with a ring kernel (inner radius of 15\arcmin\ and a width of 10\arcmin) to compute the background stellar density and standard deviation on larger scales. The significance ($\sigma$ in Table \ref{tab:all_detections}) of a pixel is then computed as the small-scale stellar density less the large-scale background density, in excess over the background standard deviation. We do not explicitly account for internal holes in the photometric coverage.

For each distance, we record a list of overdensities ordered by significance and compile detections across all distances. For locations with detections across multiple distances, we only retain the distance at which the stellar overdensity is most prominent. 
The final list of detections across the UNIONS sky is cross-matched with the coordinates of known Milky Way satellites, M31 satellites, and Local Group galaxies \citep{Pace25}. 
We then visually inspect 100 most significant candidate detections. This procedure was carried out independently in the NGC and SGC regions of UNIONS, and using the $gi$ and $ri$ combinations to probe regions of the sky not covered by the $gr$ data (See Fig. \ref{fig:unions-cov}).


\subsection{Detections}
\label{subsec:lists}

We visually assessed statistically significant overdensities by looking for good agreement between stars plotted on color-magnitude diagrams (CMDs) with an old, metal-poor isochrone. For promising candidates, we also examined the imaging directly to see if stellar overdensities were directly visible, or if imaging artifacts such as bright stars or chip gaps could be erroneously increasing stellar overdensity significance. We looked at the 100 most statistically significant overdensities in six lists, namely the lists of detections found separately in the primary NGC region and the smaller SGC region near M31, searched with each filter combination ($gr$, $gi$, \& $ri$). We make all six lists of 100 detections publicly available online (see Table 1 caption).

We identify three previously unknown, high significance candidate Milky Way satellites in the NGC region in the $gr$ search. We show the 31 most significant overdensities from this list in Table \ref{tab:all_detections} where these three candidates are the 16th (\boovi), 25th (\uii), and 31st (\uiii) most prominent detections overall, but more importantly, they are the 1st, 2nd, and 4th most prominent detections {\it not already associated with a known extragalactic system}. A single false positive is slightly more prominent than one candidate but we found that the high stellar density in this region is produced artificially by a chip gap in both the $g$ and $r$ imaging. 

We identify one additional new candidate (\draiii) found exclusively in the $gi$ NGC search with lower statistical significance, ranked as the 39th most significant detection (24th most significant when known systems are removed from the list). We visually inspected many false-positive overdensities in the $gi$ lists before noting that this candidate could be worth investigating further. The strongest evidence for its credibility as a coherent, physical stellar system comes from the identification of nearly two dozen stars with similar proper motions (See Sec. \ref{subsec:dra3}).
This candidate is found at such low relative significance in $gi$ in part due to several peculiarities of the $g$-band imaging: a) there is a tendency for $g$-band point source classification to fail and produce spatially clustered groups of false stars, b) bright, saturated star halos also erroneously produce clusters of point-sources, and c) the $g$-band imaging contains more internal holes than the highly homogeneous $r$ and $i$ imaging, resulting in rapid changes in background source density that can impact the local overdensity calculations. 

The candidates presented here exemplify the difficulty of identifying new, faint satellites in survey data; \boovi, \uii, and \uiii\ are simply the most prominent statistical overdensities not associated with known structures whereas the manual examination of several hundred candidates allowed \draiii\ to be identified by the specific patterns of its stellar population, despite low statistical significance.

This search successfully detects nearly all previously known dwarf galaxies in the Milky Way halo that fall in the UNIONS footprint in at least one pair of photometric bands at high significance. The only previously known galaxy that is challenging to detect is Bo\"otes IV \citep{Homma19}, which is unsurprising as it was discovered in a deeper survey, HSC-SSP. It is not present in the list of the top 100 most significant $gr$ detections, but it is found in the top 100 in $gi$ at rank 68, slightly lower in significance than \draiii.

In addition to the Milky Way dwarfs, the UNIONS data are sensitive to dwarfs of at least $M_V \sim -6$ at the distance of the M31 halo ($\sim750$\,kpc) and we recover several dwarf satellites of M81 down to $M_V \sim -10$. However, no additional M31 satellites or isolated Local Group members are reported. A full analysis of the dwarf galaxy detection efficiency in the UNIONS survey as a function of intrinsic satellite properties and data quality is in preparation.

\begin{deluxetable}{cccc}
\setlength{\tabcolsep}{3pt}
\tabletypesize{\footnotesize}
\tablecaption{List of most significant detections in UNIONS $gr$ photometry with new discoveries noted in {\bf bold}
\label{tab:all_detections}
}
\tablehead{
\colhead{R.A.} & \colhead{Dec.} & \colhead{$\sigma$} & \colhead{Notes}
}
\startdata
260.01210 & $+57.89850$ & 115.13 & Draco \\
227.05740 & $+67.16660$ & 103.49 & Ursa Minor \\
251.79410 & $+47.53050$ & 89.38 & NGC 6229 \\
259.16230 & $+43.20730$ & 71.83 & M92 \\
149.80380 & $+30.75250$ & 70.51 & Leo A \\
250.56990 & $+36.40420$ & 51.50 & M13 \\
162.32920 & $+51.04240$ & 34.66 & Willman 1 \\
210.66020 & $+54.28880$ & 23.22 & NGC 5447 \\
132.84620 & $+63.12840$ & 21.06 & Ursa Major II \\
184.12760 & $+69.41870$ & 20.79 & NGC 4236 \\
225.41650 & $+66.95610$ & 18.24 & Munoz 1 \\
177.67230 & $+38.85810$ & 18.02 & DDO 99 \\
112.13750 & $+69.16240$ & 17.59 & UGC 3847 \\
169.58780 & $+58.01440$ & 17.55 & NGC 3613 \\
183.84330 & $+36.34850$ & 17.06 & DDO 113 \\
235.90360 & $+39.99870$ & 15.37 & {\bf \boovi} \\
238.17580 & $+64.55850$ & 14.55 & Draco II \\
171.16680 & $+38.71830$ & 13.82 & NGC 3665 \\
179.52570 & $+55.44980$ & 12.21 & NGC 3998 \\
194.28270 & $+34.31270$ & 11.72 & Canes Venatici II \\
156.96970 & $+68.40210$ & 11.64 & IC 2574 \\
211.80700 & $+54.95820$ & 10.82 & NGC 5485 \\
185.00000 & $+75.34890$ & 10.73 & NGC 4291 \\
211.69310 & $+50.71100$ & 10.44 & NGC 5481 \\
186.11340 & $+51.15050$ & 10.40 & {\bf \uii} \\
206.74030 & $+60.93400$ & 10.26 & NGC 5308 \\
211.17790 & $+53.63930$ & 9.88 & NGC 5474 \\
191.05580 & $+32.28180$ & 9.84 & NGC 4657 \\
152.35900 & $+36.94410$ & 9.78 & {\it False Positive} \\
148.63460 & $+69.15770$ & 9.68 & M81 \\
237.17080 & $+65.98100$ & 9.66 & {\bf \uiii} \\
\ldots & \ldots & \ldots & \ldots \\
\enddata
\tablecomments{\draiii\ is not within the $gr$ footprint and therefore not present in this list. It is instead found in the $gi$ footprint.}
\tablecomments{The full tables contain the 100 most significant detections from the search in each band combination and can be found in full at the following link: \url{https://github.com/simon-et-smith/State-of-the-UNIONS-Candidates.git}}
\end{deluxetable}

\section{Measurement Methods}
\label{sec:methods}

Here, we detail the basic photometric measurements applied the four newly discovered Milky Way companions.
The final values derived for each observational and physical parameter are listed in Table \ref{tab:final-properties}.

\begin{deluxetable*}{clccccc}
\setlength{\tabcolsep}{2pt}
\tabletypesize{\footnotesize}
\tablewidth{0pt}
\tablecaption{Measured and derived properties for Milky Way satellite candidates \label{tab:final-properties}}

\tablehead{
    Property & Description & \uii & \uiii & \boovi & \draiii & Unit
}

\startdata
$\alpha_{J2000}$ & Right Ascension & 12\h24\m28\seco.2$\, \pm \, 0.3$ & 15\h48\m47\seco.2$\, \pm \, 0.3$ & 15\h43\m 39\seco.0$\, \pm \, 0.6$ & 18\h41\m4\seco$\, \pm \, 1$ & hms \\
$\delta_{J2000}$ & Declination & 51\textdegree9\arcmin8\arcsec$\pm \, 4$ & 65\textdegree58\arcmin56\arcsec$\pm \, 5$ & 40\textdegree0\arcmin5\arcsec$\pm \, 12$ & 75\textdegree40\arcmin41\arcsec$\pm \, 15$ & dms \\
$\alpha_{J2000}$ & Right Ascension & $186.117\pm0.001$ & $237.197\pm0.001$ & $235.912\pm0.002$ & $280.267^{+0.005}_{-0.004}$ & deg \\
$\delta_{J2000}$ & Declination & $51.152\pm0.001$ & $65.982\pm0.001$ & $40.002\pm0.003$ & $75.678\pm0.004$ & deg \\
$r_{\text{h,ang}}$ & Angular Half-Light Radius & $0.41^{+0.07}_{-0.09}$ & $0.5^{+0.1}_{-0.2}$ & $1.7^{+0.3}_{-0.4}$ & $1.4^{+0.4}_{-0.7}$ & \arcmin \\
$r_{\text{h,phys}}$ & Physical Half-Light Radius & $5\pm1$ & $10^{+3}_{-2}$ & $54^{+15}_{-13}$ & $51^{+30}_{-19}$ & pc \\
$\epsilon$ & Ellipticity\tablenotemark{a} & $0.28\pm0.18$ & $0.20^{+0.18}_{-0.14}$ & $0.36^{+0.15}_{-0.20}$ & $0.26^{+0.22}_{-0.18}$ & -- \\
$\theta$ & Position Angle & $123^{+24}_{-25}$ & $198^{+49}_{-77}$ & $167^{+15}_{-17}$ & $146^{+66}_{-57}$ & deg \\
$N^*$ & Number of Visible Stars\tablenotemark{b} & $24\pm5$ & $26^{+6}_{-5}$ & $69^{+12}_{-11}$ & $18^{+6}_{-5}$ & -- \\
$\tau$ & Age (Isochrone)\tablenotemark{c} & $12$ & $12$ & $12$ & $12$ & Gyr \\
\met & Metallicity (Isochrone)\tablenotemark{c} & $-2$ & $-2$ & $-2$ & $-2$ & dex \\
$D_{\odot}$ & Heliocentric Distance\tablenotemark{d} & $40\pm5$ & $65\pm5$ & $110\pm15$ & $130\pm30$ & kpc \\
$D_{\text{MW}}$ & Galactocentric Distance & $43\pm5$ & $67\pm5$ & $108\pm15$ & $132\pm30$ & kpc \\
M$_{\text{tot}}$ & Total Stellar Mass & $130^{+50}_{-40}$ & $280^{+110}_{-80}$ & $1500^{+700}_{-500}$ & $5000^{+4000}_{-2000}$ & \Msun \\
$M_V$ & Absolute $V$-band Magnitude & $+0.4^{+0.6}_{-0.9}$ & $-0.7^{+0.6}_{-0.8}$ & $-2.7^{+0.6}_{-0.5}$ & $-4.0\pm0.6$ & mag \\
$N_{\text{tot}}$ & Total Number of Stars & $400^{+200}_{-100}$ & $1000^{+400}_{-300}$ & $5000\pm2000$ & $16000^{+13000}_{-7000}$ & -- \\
$\mu_{\text{eff}}$ & Effective Surface Brightness & $26.9^{+0.7}_{-1.0}$ & $27.6^{+0.8}_{-0.9}$ & $29.0^{+0.6}_{-0.5}$ & $27.8^{+1.0}_{-0.8}$ & mag\,arcsec$^{-2}$ \\
$\mu_{\alpha}\cos{\delta}$ & Proper Motion in R.A.\tablenotemark{e} & -- & -- & $-0.3\pm0.2$ & $-0.1\pm0.1$ & mas\,year$^{-1}$ \\
$\mu_{\delta}$ & Proper Motion in Dec.\tablenotemark{e} & -- & -- & $-0.1\pm0.2$ & $0.29\pm0.08$ & mas\,year$^{-1}$ \\
$g_{15\sigma}$ & Local $g$-band $15\sigma$ depth & $24.2$ & $24.3$ & $24.7$ & $23.8$ & mag \\
$r_{15\sigma}$ & Local $r$-band $15\sigma$ depth & $24.1$ & $24.4$ & $24.5$ & $23.9$ & mag \\
$i_{15\sigma}$ & Local $i$-band $15\sigma$ depth & $23.1$ & $23.2$ & $23.1$ & $23.1$ & mag \\
$E(B-V)$ & Extinction\tablenotemark{f} & $0.015$ & $0.031$ & $0.015$ & $0.102$ & mag \\
\enddata
\tablenotetext{a}{We quote the median ellipticities here and use them for all subsequent calculations. However, for \uiii\ and \draiii, the MCMC fitting routine prefers a value of zero so we here we record their 68(95)\% upper limits as $\epsilon<$ 0.28(0.49) and 0.36(0.61), respectively.}
\tablenotetext{b}{For \draiii, we do not use $N^*$ for the mass estimation due to concerns about incompleteness. We instead manually count 14 stars along the isochrone brighter than the restrictive completeness limit of $i=22$ and model uncertainties using Poisson statistics.}
\tablenotetext{c}{We emphasize that the age and metallicity are not well measured at this time, but a 12\,Gyr, $-2$\,dex isochrone is selected to represent the stellar populations for all other calculations.}
\tablenotetext{d}{We emphasize that the distance uncertainties are estimated from visual inspection of isochrones overlaid on fundamentally sparse stellar populations. These uncertainty estimates are primarily of utility when they are propagated through to the stellar mass/absolute $V$-band magnitude measurements.}
\tablenotetext{e}{Quoted uncertainties are statistical. There is an additional systematic uncertainty of $\pm$\,0.033\,mas\,year$^{-1}$\citep{Lindegren21}.}
\tablenotetext{f}{Extinction values are the average value of the maps from \citet{Schlegel98}, re-calibrated following \citet{Schlafly11}, within a $10$\arcmin$\times10$\arcmin\ region towards each satellite.}

\end{deluxetable*}

\subsection{Age, Metallicity, and Distance}
\label{subsec:AMD}

We preface this subsection by stating that even in well-populated stellar systems, the so-called age-metallicity degeneracy limits the information that can be used to simultaneously determine age, metallicity, and by proxy, distance, from comparisons with theoretical isochrones \citep{Gallart05}. The additional challenge for these four discoveries is that we are also severely limited by the number of visible stars. 
In the brightest system, \draiii, we identify several RGB stars but our photometric depth is suspect (see Sec. \ref{subsec:dra3}) and does not reach the base of the RGB, so we do not have access to the main sequence (MS) or main sequence turn-off (MSTO).
The faintest system reported here, \uii, is closer and as such the MS is identified, but there is not a single obvious giant star so the RGB provides no information.
As such, the reported ages and metallicities are not measurements, but instead {\it assigned} values to be used for isochrone models. The measured distances are therefore explicitly dependent on how isochrone features (RGB, MS) line up with the data. When spectroscopy is obtained that can anchor the mean metallicity of each system, the distance and age will be subject to revision.

We first make photometric quality cuts at both the bright and faint ends. We estimate the global saturation limits of each band using $\sim 2$ million sources cross-matched to the Pan-STARRS 3$\pi$ survey \citep{Chambers16}, finding that bright sources in UNIONS photometry deviate from the shallower 3$\pi$ magnitudes at $g \sim 17$, $r \sim 17.5$, and $i \sim 14.5$, respectively (See App. \ref{app:sat} for further details). We find from a comparison with deeper HSC-SSP data that the $r$-band is $94\%$ complete at 15$\sigma$ while completeness drops to $78\%$ at 10$\sigma$. We therefore use make a faint-end cut at the median 15$\sigma$ point source depth in a 2\textdegree\ radius circle centered on each candidate measured in each band to ensure a high degree of completeness for the sake of estimating structural parameters and total magnitudes of each satellite. We refer to these faint-end cuts as `local depths' and list them in Table \ref{tab:final-properties}.

We visually inspect a set of PARSEC isochrones overlaid on CMDs constructed with UNIONS photometry selected within a 2 arcminute diameter aperture of the initial detection coordinates.
For each candidate, we visually examine a grid of isochrones with ages of 5 to 13 \,Gyr and metallicities of $-1$ to $-2.2$\,dex that are shifted to the initial distance estimate from the search routine, overlaid on de-extincted photometry. We also modify the distance in small (5\,kpc) steps to best match the photometry. In general, isochrones older than 10\,Gyr that are more metal-poor than $-1.5$\,dex provide good fits to the data, but the degeneracy between age, metallicity, and distance precludes us from achieving confident estimates.

For the remainder of the analysis, we select an isochrone of 12\,Gyr, [Fe/H] = $-2$ to model the underlying stellar population which represents the basis for estimation structural parameters, identifying member stars, and measuring the stellar mass. Future revisions to the isochrone model will therefore necessitate updates to all of these properties. Fixing the isochrone properties now allow us to move forward with the analysis. 
We report first pass distance estimates with 10--25\% uncertainties, which we justify for each target in Sec. \ref{sec:results} based on specific features in the photometry in relation to the representative isochrone.






\begin{deluxetable*}{llccccccc}
\label{tab:bhb}
\tabletypesize{\small}
\tablewidth{0pt}
\tablecaption{Candidate Blue Horizontal Branch Stars for \boovi\ and \draiii}

\tablehead{
    Galaxy & {\it Gaia} ID & R.A. & Dec. & $P_{\mathrm{sat}}$ & $g_0$ & $(g-r)_0$ & $M_g$ & $D_\odot$ \\
    -- & -- & deg & deg & -- & mag & mag & mag & kpc 
}

\startdata
\hline
\boovi & 1377856157443591808 & 235.94059 & 39.93902 & 0.56 & $20.61\pm0.03$ & $-0.18\pm0.05$ & $0.51\pm0.20$ & $105\pm10$ \\
\draiii & 2268520812859202048 & 280.59427 & 75.66594 & 0.63 & $20.24\pm0.03$\tablenotemark{a} & $-0.09\pm0.06$\tablenotemark{a} & $0.46\pm0.16$ & $90\pm7$ \\
\enddata
\tablenotetext{a}{\draiii\ is outside of the SDSS survey area, so the values of $g_0$ and $(g-r)_0$ have been obtained by converting from Pan-STARRS 3$\pi$ photometry following \citet{Thomas18}.}
\end{deluxetable*}

\subsubsection{Blue Horizontal Branch Stars}
\label{subsubsec:bhb}

Where possible, we use blue horizontal branch (BHB) stars to compute the distance to each satellite and compare to the by-eye isochrone estimates. We use the calibration given by Equation (7) in \citet{Deason11} to obtain the absolute $g$-band magnitude ($M_g$) of BHB stars as a function of $(g-r)_0$ in SDSS AB magnitudes \citep{Abazajian09}. Stars must have $-0.25 < (g-r)_{0,\mathrm{SDSS}} < 0$ to be well described by this calibration.
Some stars considered in this analysis lie outside the SDSS footprint, in which case we use Pan-STARRS 3$\pi$ photometry \citep{Chambers16} and convert to SDSS magnitudes using Equation (6) from \citet{Thomas18}, which is explicitly valid for BHB stars. We then obtain the distance modulus, and subsequent heliocentric distance, for each star using their measured apparent and estimated absolute $g$-band magnitudes. We also include a generous uncertainty on $M_g$ of 0.15\,mag based on Fig. 4 from \citet{Deason11}, which is added in quadrature to the photometric uncertainties of individual stars.

We identify one potential BHB star in \boovi\ and \draiii\, but none in \uii\ or \uiii.
We find that the BHB-based distance estimates are consistent with the first-pass isochrone-based distances estimates, but have much smaller uncertainties than the isochrone-based estimates done by eye. The results of this exercise can be found in Table \ref{tab:bhb}.


\subsection{Structural Parameters}
\label{subsec:structure}

We compute structural parameters following the methodology of \citet{Martin08, Martin16-pds} by assuming that the tangent-plane projected distribution of individual stars toward each candidate is well described by an elliptical, exponential surface density profile superimposed on a constant surface density field of contaminant non-members in the form of Milky Way disk and halo stars as well as faint background galaxies that were not removed by the star-galaxy separation. The profile, $\rho_{\text{dwarf}}(r)$, is parameterized by the centroid of the profile ($x_0, y_0$), the ellipticity $\epsilon$ (defined as $\epsilon = 1 - b/a$ where $b/a$ is the minor-to-major-axis ratio of the model), the position angle of the major axis $\theta$ (defined East of North), the half-light radius (which is the length of the semi-major axis \rh), and the number of observed stars $N^*$ in the color-magnitude selection region. The model is written as 

\begin{equation}
	\rho_{\text{dwarf}}(r) = \frac{1.68^2}{2\pi\rhm^2(1-\epsilon)}N^*\exp\bigg(\frac{-1.68r}{\rhm}\bigg)
\end{equation}

\noindent where $r$, the elliptical radius, is related to the tangent-plane projected sky coordinates ($x$, $y$) by

\begin{eqnarray}
    \nonumber r = \Bigg\{ \bigg[\frac{1}{1-\epsilon} \Big((x-x_0)\cos\theta - (y-y_0)\sin\theta\Big) \bigg]^2 \\ + \bigg[ (x-x_0)\sin\theta + (y-y_0)\cos\theta \bigg]^2\Bigg\}^{\frac{1}{2}}.
\end{eqnarray}

\noindent We assume that the background stellar density is uniform, which is reasonable on the scale of arcminutes up to a degree or so. The background, $\Sigma_{\text{b}}$, is calculated as

\begin{equation}
    \Sigma_{\text{b}} = \frac{n - N^*}{A},
\end{equation}

\noindent where $n$ is the total number of stars in the field of view and $A$ is the total area, normalizing the background density with respect to the selected region. We combine the elliptical, exponential surface density model with the uniform background term to construct our posterior distribution function,

\begin{equation}
    \rho_{\text{model}}(r) = \rho_{\text{dwarf}}(r) + \Sigma_{\text{b}}.
\end{equation}

For each candidate, we perform a selection around the representative isochrone that is identical to the filtering step in the search algorithm, i.e. stars are selected within 0.1\,mag including typical photometric uncertainties added in quadrature. We also apply the bright and faint magnitude cuts to obtain a clean sample of potential member stars. We only use the MS and RGB regions of the isochrone to perform this selection. We ignore the horizontal branch (HB) as its exact morphology is complex \citep[See][and references therein]{Gallart05} and, due to the rarity of these evolved stars in faint stellar systems, we conclude that selecting this region of the CMD will decrease signal-to-noise.

We determine the best-fitting structural parameters by using the Affine-invariant Markov Chain Monte Carlo Ensemble sampler \texttt{emcee} \citep{Foreman13} to sample the posterior distribution. We use 64 walkers which each take 5,000 steps, with the first 1,000 thrown out as burn-in. Flat priors are applied to each parameter and are listed in Table \ref{tab:priors}. Open and closed intervals are indicated with round and square brackets respectively. Most priors are identical for each system. Note that $\theta$ is constrained to a 180\textdegree\ interval where the starting point was adjusted after the first run to ensure that the resulting median was roughly in the middle of the interval. Additionally, $N_{\rm max}$ refers to the total number of stars selected for the fitting routine per system.
The median values and 16th/84th quantiles of the posterior distribution for each parameter are given in Table \ref{tab:final-properties} and show corner plots in App. \ref{app:corner}. Note that for \uiii\ and \draiii, the preferred value of the ellipticity is zero so we also record the 68th(95th) quantiles of the chains produced from the MCMC. However, we use to still use the median ellipticity values for the sake of estimating member stars in {\it Gaia} (Sec. \ref{subsec:members}) and surface brightness (Sec. \ref{subsec:mass}).

\begin{deluxetable}{lcccc}
\tabletypesize{\footnotesize}
\tablecaption{Flat priors for each parameter in the MCMC analysis, per candidate \label{tab:priors}}

\tablehead{
     & \uii & \uiii & \boovi & \draiii
}

\startdata
$x_0$ & [$-15$, 15] & [$-15$, 15] & [$-15$, 15] & [$-15$, 15] \\
$y_0$ & [$-15$, 15] & [$-15$, 15] & [$-15$, 15] & [$-15$, 15] \\
\rh & (0, 15) & (0, 15) & (0, 15) & (0, 15) \\
$\epsilon$ & [0, 1) & [0, 1) & [0, 1) & [0, 1) \\
$\theta$ & [30, 210] & [90, 270] & [90, 270] & [60, 240] \\
$N^*$ & [0, $N_{\rm max}$] & [0, $N_{\rm max}$] & [0, $N_{\rm max}$] & [0, $N_{\rm max}$] \\
\enddata

\end{deluxetable}

\subsection{Stellar Membership and Proper Motions}
\label{subsec:members}

We leverage the all-sky proper motion measurements from the Third Data Release (DR3) from {\it Gaia} \citep{Gaia16,Gaia23} to search for member stars in each satellite and determine a systemic proper motion where possible. We employ the methodology developed by \citet{McVenn20a, McVenn20b} and \citet{Jensen24} and direct readers to those works for a full description of the procedure. In brief, the method uses the representative isochrone, estimated distance, and structural parameters to estimate membership likelihoods for all stars with full astrometric data \citep[][]{Lindegren21}, compared to the Milky Way foreground population, and compute a systemic proper motion for each satellite. We successfully identify several candidate member stars and a systemic proper motion for \boovi\ and \draiii, but we do not find any likely members for \uii\ and \uiii, reflecting their complete lack of giant stars.





\subsection{Stellar Mass and Total V-band Magnitude}
\label{subsec:mass}

We compute the total stellar mass and absolute $V$-band magnitude of each satellite using the same approach presented in \citet{Smith24, Smith25}, which is based on the methodologies developed in \citet{Martin08, Martin16-pds}. We model each satellite as a simple stellar population (SSP) corresponding to their representative isochrone that obeys an underlying Kroupa initial mass function \citep{Kroupa01} down to the hydrogen burning limit (M$_* = 0.08$\Msun). We generate 10,000 synthetic SSPs where each SSP is assigned a random distance drawn from a normal distribution centered on $D_{\odot}$ with a $\sigma$ equal to the measured uncertainty, and a random value of $N^*$ drawn from the chains generated by the structural parameters MCMC fitting routine. In this way, uncertainty associated with estimating $D_{\odot}$ and $N^*$ are propagated through to the final distributions of total stellar mass and total $M_V$.

In the structural parameter fitting procedure, $N^*$ represents the number of visible stars estimated to be in the stellar overdensity. Recall, we make a bright cut at the saturation limits, a faint cut at the local 15$\sigma$ point source depths, and only consider the MS and RGB when selecting the sample of putative member stars. Therefore, in order to generate a single synthetic SSP that emulates the observations, we stochastically sample stellar masses from the IMF and check whether the apparent magnitude of that star at the drawn distance $D_{\odot}$ falls in the magnitude regime defined by the extinction-corrected $15\sigma$ depth in both bands at the faint end, and the saturation limits at the bright end. We proceed until we have accumulated $N$ stars equal to the $N^*$ drawn for that particular SSP, at which point we record the total stellar mass of all stars sampled. 
We then use a PARSEC isochrone computed for the Johnson's photometric system \citep[][]{Bessel90, MaizAppelaniz06} with the same age and metallicity as the representative isochrone to obtain absolute $V$-band magnitudes for each star, and sum the equivalent luminosities to obtain the total absolute $V$-band magnitude for the SSP. We compute the effective surface brightness (the average surface brightness within the elliptical \rh) for each synthetic population.

Finally, we record the systemic values of total stellar mass, total luminosity, total absolute $V$-band magnitude, and effective surface brightness as the median from 10,000 synthetic SSPs, with upper and lower uncertainties quoted as the 16th and 84th quantiles of the ensemble. Note that these derived stellar masses do not account for white dwarfs, neutron stars, or black holes that may contribute to the total baryonic mass budget. Stellar evolution considerations paired with detailed orbital modeling and the impact of mass loss may be necessary to correctly estimate the amount of non-dark matter mass in a given system \citep{Devlin25, Rostami-Shirazi25}.

\begin{figure*}
    \centering
    \includegraphics[width=0.7\linewidth]{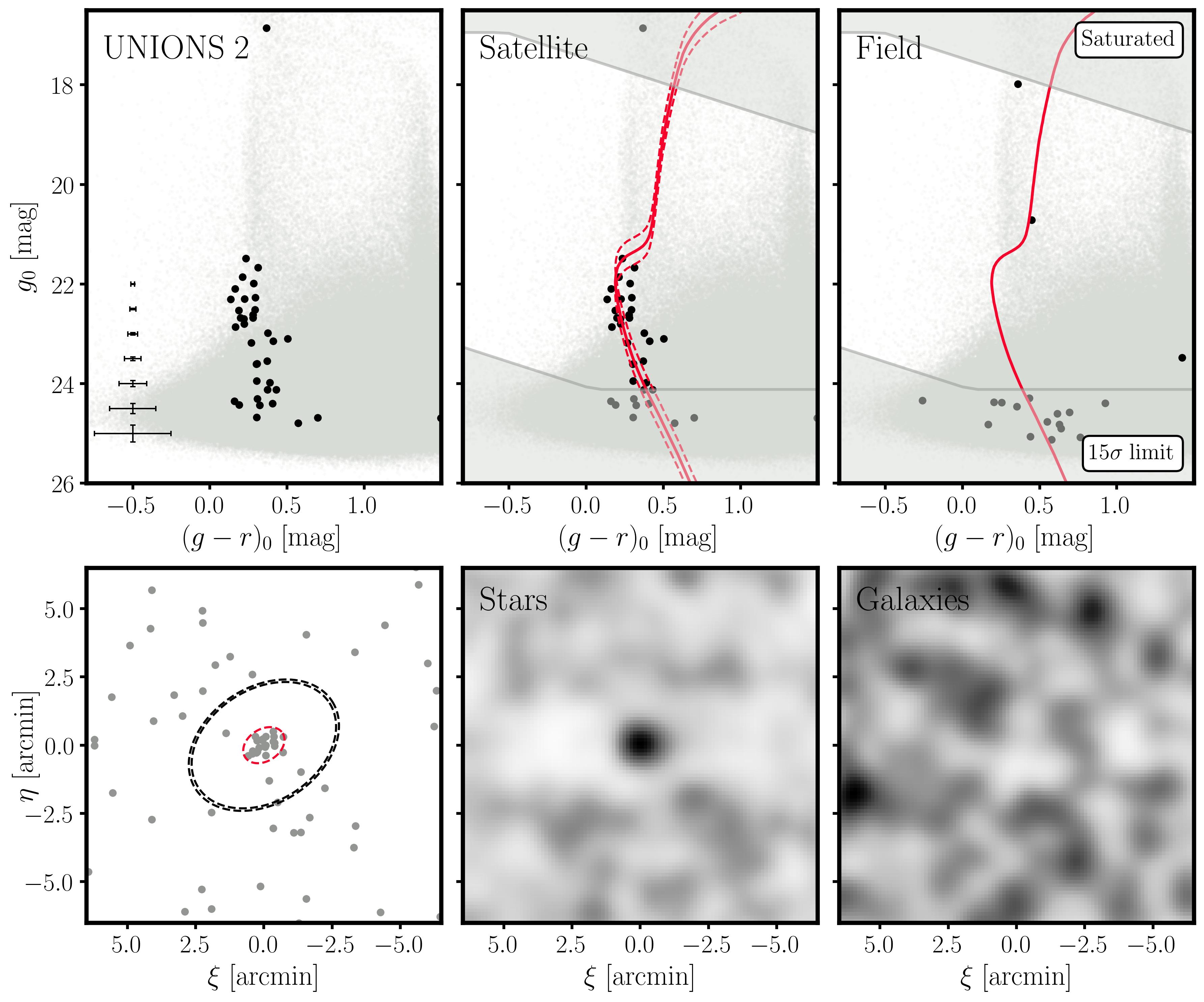}
    \caption{Summary plot for the discovery of UNIONS 2. 
    {\it Upper Row}: Three panels showing CMDs for the candidate satellite and a reference field. Light grey points in all three CMDs shows all stars in a 2 degree circle around the centroid. {\it Left}: Black markers show all stars contained within 2 elliptical \rh\ of the centroid. Black crosses on the left side indicate the median photometric errors as a function of magnitude. {\it Center}: Same as the upper left, but the best-fitting isochrone for \uii\ is overlaid as a solid red line, while dashed red lines show the same isochrone shifted to the estimated distance uncertainty of 5\,kpc. The greyed out regions at the top and bottom indicate where saturation and the 15$\sigma$ magnitude limits render photometry unreliable. {\it Right}: A CMD reference field, selected as an elliptical annulus equal in area to the 2\rh\ central ellipse.  
    {\it Lower Row}: Three spatial plots showing tangent-plane projected sources in an $13\times13$ arcminute region around the centroid of UNIONS 2. 
    {\it Left}: Grey markers denote all stars selected as within 0.1\,mag of the isochrone, including photometric uncertainties added in quadrature. The red dashed ellipse demarcates 2\rh\ and encompasses all stars shown on the upper left and central panel CMDs while the concentric dashed black ellipses show the selection annulus for the upper right panel CMD, where the annulus begins at 7\rh. {\it Center}: Binned and smoothed distribution of all stellar sources in the region {\it without} any membership selection. {\it Right}: Distribution of all non-point sources (i.e. galaxies) in the same patch of sky.}
    \label{fig:sum-u2}
\end{figure*}

\begin{figure*}
    \centering
    \includegraphics[width=0.7\linewidth]{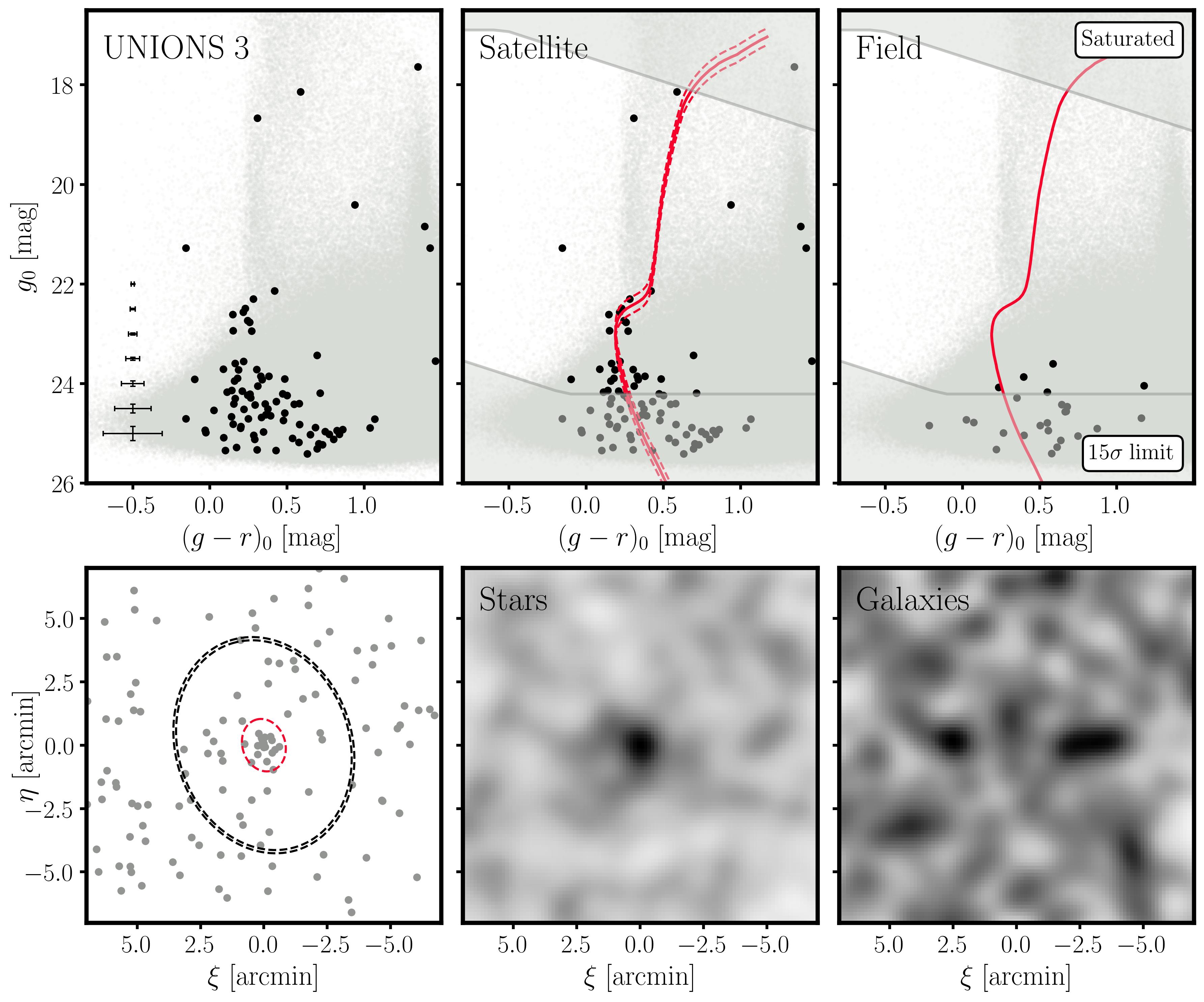}
    \caption{Summary plot for the discovery of UNIONS 3. The general description for all panels is identical to that of Fig. \ref{fig:sum-u2}. In this case, the spatial region extends over an area of $14\times14$ arcminutes and the reference field selected by the elliptical annulus that begins at 8\rh.}
    \label{fig:sum-u3}
\end{figure*}

\begin{figure*}
    \centering
    \includegraphics[width=0.7\linewidth]{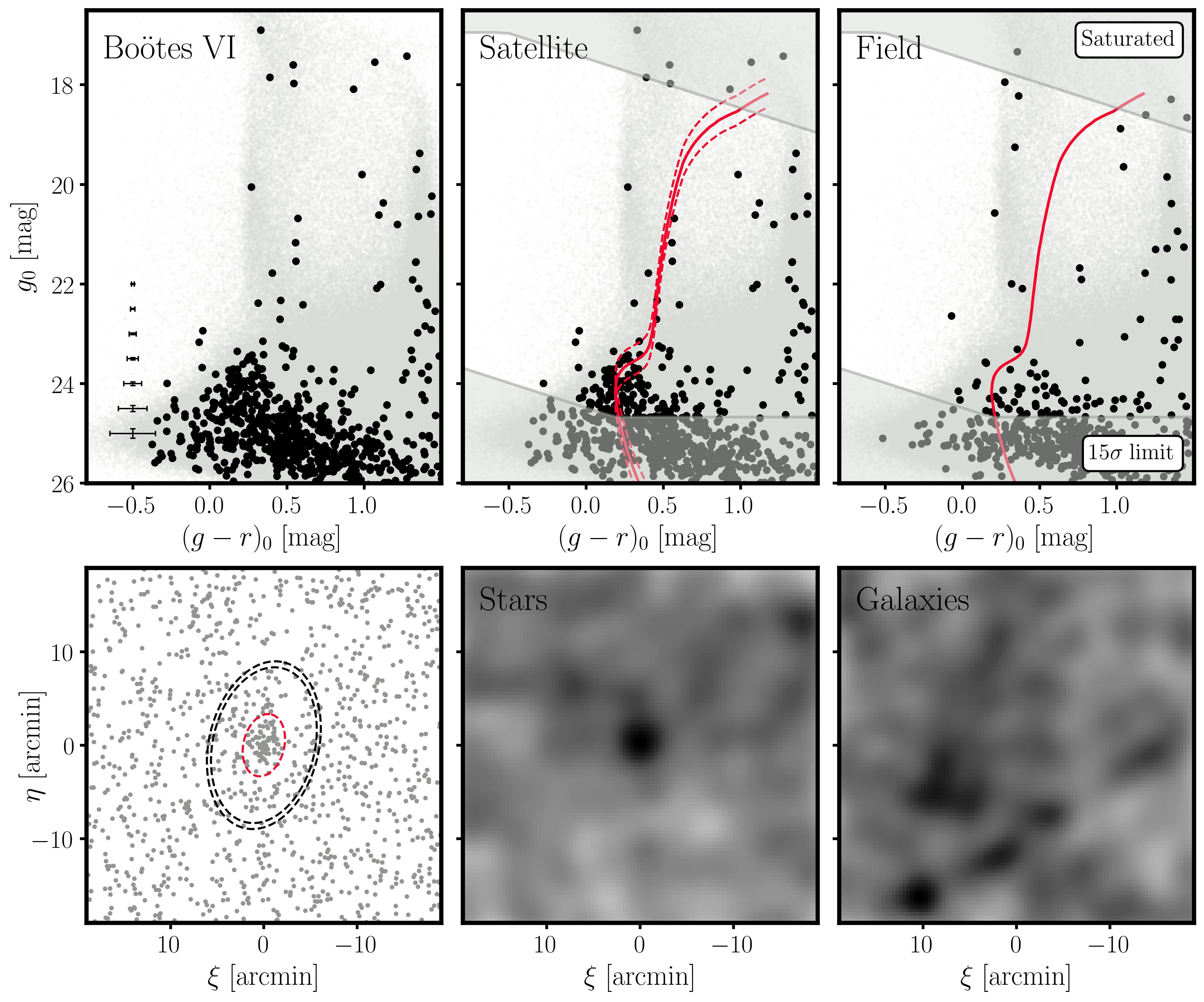}
    \caption{Summary plot for the discovery of Bo\"otes VI. The general description for all panels is identical to that of Fig. \ref{fig:sum-u2}. In this case, the spatial region extends over an area of $38\times38$ arcminutes. We also show stars in an inner region selected by the ellipse at 2\rh, and the reference field selected by the elliptical annulus that begins at 5\rh. The ellipticity is the median from the MCMC-based parameter estimate.}
    \label{fig:sum-boo6}
\end{figure*}

\begin{figure*}
    \centering
    \includegraphics[width=0.7\linewidth]{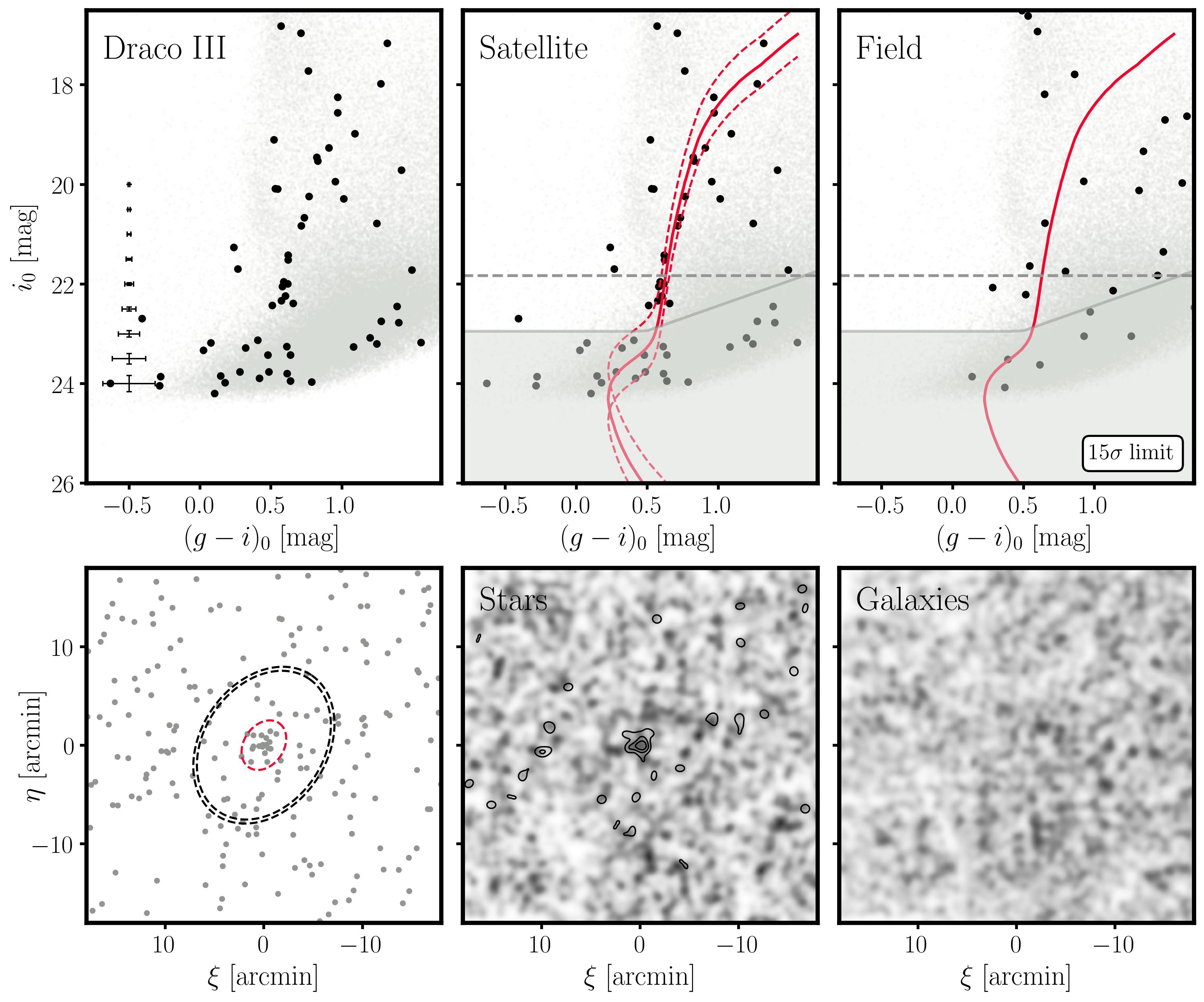}
    \caption{Summary plot for the discovery of Draco III. The general description for all panels is identical to that of Fig. \ref{fig:sum-u2}. \draiii\ is presented and investigated in $gi$ photometry because of a chip gap in the $r$-band data (see Sec. \ref{subsec:dra3}).
    In this case, the spatial region extends over an area of $36\times36$ arcminutes. We also show stars in an inner region selected by the red ellipse at 2\rh, and the reference field selected by the black elliptical annulus begins at 6\rh. Additionally, there is significant photometric source loss below $i \sim 22$\,mag (see App. \ref{app:iband}), which is denoted by a horizontal dashed grey line in the upper central and right panels. The smoothed distribution of all stellar sources does not show a particularly strong overdensity, so we also overlay the smoothed distribution when just member stars are selected, shown with black contours. We note that despite the low detection significance, we find 21 co-moving stars in {\it Gaia} proper motions, strongly suggesting that \draiii\ is a coherent stellar structure.}
    \label{fig:sum-dra3}
\end{figure*}

\section{Results}
\label{sec:results}
We now review the measured properties of each satellite and note any specific changes made to the measurement procedures for a given system. We provide an initial classification of each system given available information.
We show a summary plot for each satellite containing structural and photometric properties in Figs. \ref{fig:sum-u2} (\uii), \ref{fig:sum-u3} (\uiii), \ref{fig:sum-boo6} (\boovi), \& \ref{fig:sum-dra3} (\draiii).


\subsection{\uii}
\label{subsec:u2}
UNIONS 2 is the least massive system detected in this work.
Composed of 400$^{+200}_{-100}$ total stars (with $\sim 24$ observed in UNIONS), its total stellar mass is $120^{+40}_{-30}$\Msun\ and the median $M_V$ of 10,000 representative mock stellar populations is $+0.4^{+0.6}_{-0.9}$\,mag. It is also compact ($5\pm1$\,pc) and relatively nearby ($40\pm5$\,kpc). The stellar population is strikingly sparse, with no RGB stars being found within 2\rh\ and the brightest candidate member star having observed magnitudes of $g = 21.5$ and $r = 21.3$ (see Fig. \ref{fig:sum-u2}). The additional proper motion analysis did not identify any likely members in the vicinity, confirming the complete lack of giant stars which induces significant challenges in fitting an isochrone and determining a distance. The only feature that suggests UNIONS 2 may be an old, metal-poor stellar population is that the putative MS does not continue to brighter magnitudes. We interpret this as the turn-off, and conclude that there are no stars currently in the subgiant branch (SGB) or RGB phases of stellar evolution. 

We estimate the distance to be 40\,kpc based on the brightest candidate member, assuming it to be at the MSTO. We set symmetric, 5\,kpc uncertainties that bracket the $g = 21.5$ star in question should it be just below the MSTO and a fainter star which could be on the SGB. Distance uncertainties are shown as dashed isochrones above and below the best estimate in the upper central panel in Fig. \ref{fig:sum-u2}.

\uii\ lies among the faint, compact satellites whose natures remain ambiguous in the absence of robust dynamical mass measurements. As Milky Way satellite dwarf galaxies are commonly named after the constellation in which they are found, we opt to reserve this naming convention for high likelihood candidate UFDs. The nature of \uii\ is ambiguous at the time of discover so we name it after the UNIONS survey. At the time of discovery, UNIONS 2 is tied as the fourth faintest Milky Way satellite known with $M_V = +0.4$ where DELVE 5 \citep{Cerny23-six} is also +0.4\,mag, and only Kim 3 \citep[+0.7,][]{Kim16}, DELVE 7 \citep[+1.2,][]{Tan25-three}, and Ursa Major III/UNIONS 1 \citep[+2.2,][]{Smith24} are fainter. 

\subsection{\uiii}
\label{subsec:u3}
\uiii\ is roughly twice as massive as UNIONS 2, with an estimated stellar mass of $280^{+110}_{-80}$\Msun\ and $M_V = -0.7^{+0.6}_{-0.8}$\,mag. Its stellar population contains a potential RGB star near the base of the branch, as well as a handful of possible SGB and MSTO stars. The distance is also challenging to determine but the initial estimate is 65\,kpc.
The lower limit is set where the SGB sweeps across the putative end of the MS, and the upper limit is set to still intersect with the star at the base of the RGB resulting in a distance estimate of $65\pm5$\,kpc, as seen in Fig. \ref{fig:sum-u3}. There is also a small gap in the stellar sequence below the MSTO. This may simply be due to the stochastic nature of low mass star formation as gaps of this nature have been identified before in other faint Milky Way satellites \citep[e.g.][]{Smith24}.

The membership detection procedure found no stars measured by {\it Gaia} to be consistent with the isochrone at the distance of UNIONS 3. The bright star at $g_0 \sim 18.5$ and $(g-r)_0 \sim 0.5$ on the central CMD in the top row of Fig. \ref{fig:sum-u3} is well measured in {\it Gaia}, but its parallax implies a distance of 3.1\,kpc and therefore is assumed to be a non-member when performing isochrone-based distance estimation. 


With a physical size of 10\,pc, \uiii\ sits at an interesting region of the size-luminosity plane as it is more compact than all confirmed UFDs, but larger than the majority of ambiguous systems.
We consider \uiii\ an ambiguous system as there is no clear evidence {\it either for or against} the presence of dark matter. Again, we chose to reserve constellation-based names for systems that are likely dwarf galaxies and therefore name this satellite after the UNIONS survey. 
Obtaining well measured spectroscopy to determine the stellar dynamics and chemistry of \uiii\ presents a significant observational challenge as the brightest candidate member star is fainter than $g = 22$, and a depth of more than $g = 23$ is required to obtain a sample of $\sim10$ stars.  

Interestingly, \uiii\ is very near to the low luminosity Milky Way satellite Draco 2 ($M_V = -0.8$), at an angular separation of only 1.5\,deg. However, Draco 2 is 21.5\,kpc from the Milky Way \citep{Longeard18}, making the physical separation more than 40\,kpc between the two satellites. Without kinematic information, there is no clear indication whether \uiii\ and Draco 2 are associated or if their close angular separation is purely coincidental.

\subsection{\boovi}
\label{subsec:boo6}
\boovi\ has a well populated MS (fainter than $g_0 \sim 23.5$), along with six likely RGB stars and an additional handful of SGB stars within 2\rh\ that define the old, metal-poor stellar population, as seen in Fig.  \ref{fig:sum-boo6}. These features allow us to estimate the heliocentric distance as $110\pm15$\,kpc. The lower limit is set by the top of the MSTO, while the upper limit is informed by a cluster of potential members at $(g-r)_0 = 0.5$ at the base of the RGB. The stellar density profile of \boovi\ is has a well defined ellipticity of $\epsilon = 0.36^{+0.15}_{-0.20}$.
For \boovi, $M_V = -2.7^{+0.6}_{-0.5}$ and a total stellar mass of roughly 1500\Msun, which is particularly faint relative to its physical half-light radius of $54^{+15}_{-13}$\,pc. Its effective surface brightness is $29.0^{+0.6}_{-0.5}$\,mag\,arcsec$^{-2}$, making it extremely similar in properties to the recently discovered Aquarius III UFD \citep{Cerny25}.
No confirmed Milky Way GC is more extended than 30\,pc \citep[Palomar 14,][]{Munoz18}, so we assert that \boovi\ is likely to be a galaxy and name it after the constellation of Bo\"otes, within which it is found.

\begin{figure*}
    \centering
    \includegraphics[width=0.95\linewidth]{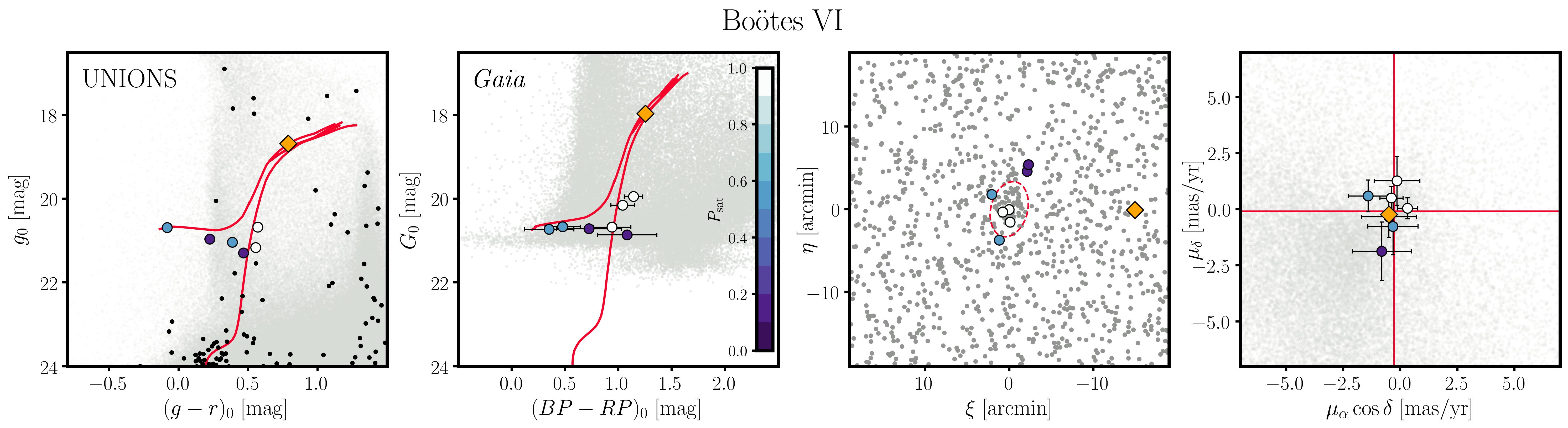}
    \includegraphics[width=0.95\linewidth]{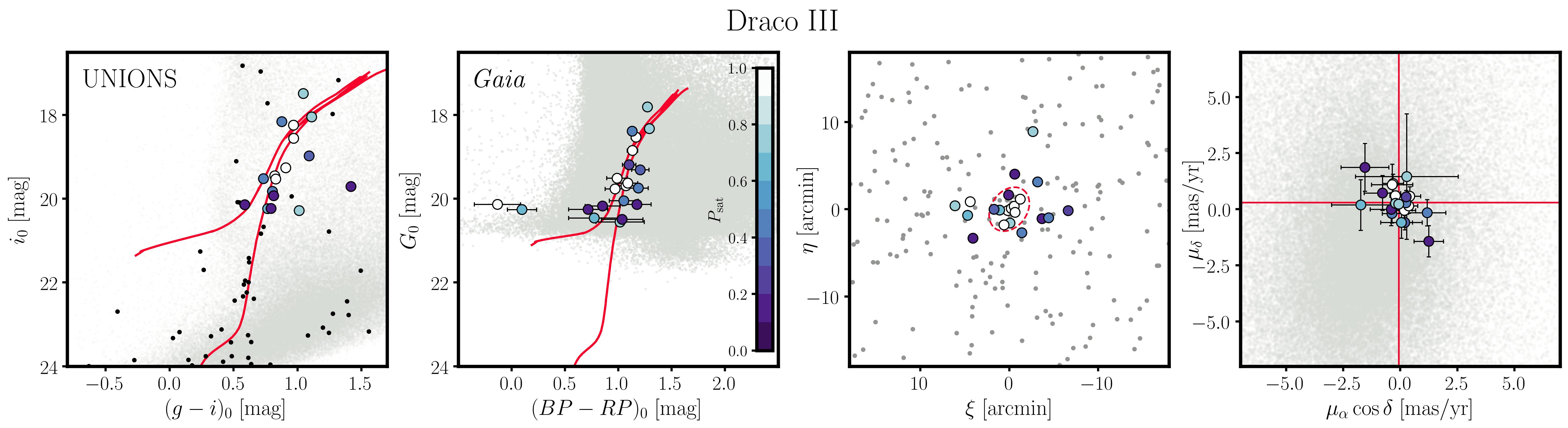}
    \caption{Photometric, spatial, and proper motion plots produced for the determination of systemic proper motions and member stars for Bo\"otes VI ({\it Upper Row}) and Draco III ({\it Lower Row}), where each column shows equivalent information for each satellite. {\it First Column}: CMD of UNIONS photometry where grey markers are the same as in the `Satellite' CMDs from the summary panels, i.e. all stars within 2\rh. The representative isochrone is overlaid in red. The circular markers show all stars with $P_{\mathrm{sat}} > 0.1$, regardless of distance to the centroid, colored by $P_{\mathrm{sat}}$. We make special note of the orange diamond marker in the upper panel, which is the possible distant member in \boovi\ that also appears as an orange diamond in all subsequent panels. 
    {\it Second Column}: CMD in {\it Gaia} photometric system. Grey points are all those within 2 degrees of the satellite. The same isochrone is shown in the {\it Gaia} bands. The color bar indicates the computed $P_{\mathrm{sat}}$ for all overlaid members, and black errorbars show individual photometric uncertainties. Note that there are one and four stars present in the {\it Gaia} data that are missing in the {\it Left} UNIONS panels for \boovi\ and \draiii, respectively, due to source detection failures in the $g$-band (See App. \ref{app:missing}).
    {\it Third Column}: Spatial plot for the same region as shown in the summary plots. The dashed red ellipse denotes 2\rh. {\it Fourth Column}: Proper motions for all suspected members (circles) and all sources within 2 degrees of the satellite (grey points). Black errorbars are individual proper motion uncertainties. The intersecting red lines indicate the measured systemic proper motion of the satellite.}
    \label{fig:cmd-gaia}
\end{figure*}

We identify several likely member stars in \boovi. We overplot all stars with $P_{\mathrm{sat}} > 0.1$ on the CMDs in Fig. \ref{fig:cmd-gaia}. \citet{Jensen24} assessed the performance of  $P_{\mathrm{sat}}$ by assuming radial velocities as the ground truth for membership and found that stars with $P_{\mathrm{sat}} > 0.2$ have a greater than 50\% likelihood of being consistent with the systemic velocity of a given satellite, while many stars with $0.1 <P_{\mathrm{sat}} < 0.2$ still often prove to be members upon radial velocity follow-up. We therefore report five high likelihood members with $P_{\mathrm{sat}} > 0.2$ and two additional medium significance candidates. We also note that one high likelihood member ({\it Gaia} ID ending in 4256) is not detected in UNIONS $g$ where it appears that a nearby, saturated star has interfered with its detection (see App. \ref{app:missing}), so it does not appear in the upper leftmost panel of Fig. \ref{fig:cmd-gaia}.

In addition to the high likelihood members, we identify an interesting star with $P_{\mathrm{sat}} \sim 0.002$ (the 8th likeliest member) that sits 13.6 elliptical \rh\ ($r_{\rm ell}$), or $14.8$\,arcminutes, from the centroid of \boovi\ along the minor axis of the system, where this large separation factors into its very low membership probability in our framework. However, it lies directly on the upper RGB and has a well measured proper motion that is consistent with the systemic value (shown as an orange diamond in Fig. \ref{fig:cmd-gaia}), so it may in fact be gravitationally bound to \boovi.
Member stars at large physical separations from their hosts have garnered significant attention in recent years. The identification of distant members through {\it Gaia} proper motions \citep{Jensen24} and RR Lyrae searches \citep{Tau24} have demonstrated that extended halos are not uncommon around UFDs, and focused studies have found that plausible explanations include tidal disruption \citep{Waller23, Ou24}, bursty star formation feedback \citep{Chiti21, Waller23}, and dwarf-dwarf mergers \citep{Chiti21, Waller23, Jensen24}. Simulations have also demonstrated that dwarf-dwarf mergers can deposit stars on long-term, high-energy around UFDs while offering insight into the merger timescales \citep{Goater24} and the nature of dark matter \citep{Deason22}. 
Failure to account for stars at large radii can also lead to underestimated satellite luminosities and stellar masses \citep{Andersson25, Riley25, Wheeler25}.
Member stars at similarly large radii have only been found in a select few galaxies with similar $M_V$ to \boovi. \citet{Tau24} find probable RR Lyrae members beyond 10\rh\ in Pegasus IV ($-4.3$), Tucana II ($-3.8$) and Tucana III ($-1.3$), although Tucana III may be tidally disrupting \citep{Drlica15, Simon17}.
\citet{Jensen24} use {\it Gaia} proper motions to identify likely members out to $\sim 10$\rh\ in Tucana II and Tucana III, as well as in Grus II ($-3.5$) and the extremely faint likely UFD Draco 2 ($-0.8$). Tucana II also has a spectroscopically confirmed member at $\sim9$\rh\ \citep{Chiti21}.
Due to the rarity of this phenomenon, we identify this possible distant member of \boovi\ as a high priority target in future spectroscopic studies, in addition to the 7 other members identified closer to the satellite centroid.

Following Sec. \ref{subsubsec:bhb}, we identify one potential BHB star among the members ({\it Gaia} ID ending in 1808) and compute its heliocentric distance to be $105\pm10$ (See Table \ref{tab:bhb}). This measurement is in good agreement with the isochrone-based distance estimate of $110\pm15$\,kpc.

\subsection{\draiii}
\label{subsec:dra3}
The discovery of \draiii\ is particularly serendipitous due to its placement on the sky. It is on the very edge of the nominal UNIONS footprint, and while there is some $r$-band imaging in the area, the centroid of \draiii\ falls in a horizontal gap between CFIS $r$-band exposures. It was instead detected in the combined $gi$ imaging (Fig. \ref{fig:sum-dra3}), where both the $g$- and $i$-bands imaging fortuitously had full spatial coverage in that region. \draiii\ is at a low galactic latitude ($b = 26.9$\,deg) and therefore has a significant galactic extinction along the line of sight of $E(B-V) = 0.118$, compared to the extinction for the three other satellites presented in this work (all with $E(B-V) < 0.04$).

Examining the CMD in Fig. \ref{fig:sum-dra3}, we identify a relatively well populated RGB, but it does not appear that we robustly detect the MSTO. Using the representative isochrone of 12\,Gyr and $-2$\,dex, we find that distances ranging from 100 to 160\,kpc provide reasonably good fits to the upper portion of the RGB, but there are no clear member stars that are red enough to use the tip of the RGB for distance estimation. We estimate a distance of $130$\,kpc with a generous uncertainty of 30\,kpc.

Despite such a sparse CMD and a moderate spatial overdensity, we identify 18 high likelihood members ($P_{\mathrm{sat}} > 0.2$) and three additional medium probability members co-moving on the sky. This is the most compelling evidence that \draiii\ is a real, coherent stellar system.
Four of the 21 total members do not have associated $g$-band photometry due to the presence of bright stars or manmade satellite tracks in the immediate vicinity of each member that disrupted source detection (see App. \ref{app:missing}), which is why the UNIONS CMD has fewer stars than the corresponding {\it Gaia} CMD in the bottom row of Fig. \ref{fig:cmd-gaia}.

\draiii\ is challenging to measure in detail due to an additional issue we found in the $i$-band. The source detection appears to deviate from expectations such that at magnitudes fainter than $i = 22$\,mag, source loss becomes very prominent (See App. \ref{app:iband}) despite the 15$\sigma$ point source depth being 23.1\,mag in this region. The spatial overdensity seen in Fig. \ref{fig:sum-dra3} is heavily reduced if we only retain stars brighter than $i=22$ to try and mitigate the effects of this unusual stellar incompleteness, so much so that the structural parameter fitting code is unable to converge. We therefore chose to select stars all the way down to the $15\sigma$ point source depth limits ($g=23.8$ and $i=23.1$), like we do for all other satellites, in order to acquire enough stars to obtain a viable estimation of the structural parameters. 

However, the spatial distribution of the 21 possible member stars immediately highlights how incompleteness may be impacting our measurements.
Note in Table \ref{tab:members} that there are 10 members within 2\rh\ with the other 11 members situated outside of this boundary. In principle, 85\% of stars in the system should be contained within 2\rh\ of the elliptical, exponential profile used to model the system. This excess of outer stars could indicate that the half-light radius of \draiii\ is not well measured due to the completeness issues in the existing photometry, and the true \rh\ may be larger. Alternately, this could be a signal the presence of a diffuse stellar halo, possibly due to a dynamical disturbance (e.g. tidal disruption or a dwarf-dwarf merger). Identifying a possible excess of stars in the outer regions of \draiii\ is strong motivation for a deep photometric follow-up of \draiii\ to robustly establish its structural parameters.

The incompleteness has further implications for the measurement of $N^*$, which parametrizes the number of visible member stars in the structural fitting code. The chains produced by the MCMC-fitting routine for $N^*$ are directly used to compute the total stellar mass and total absolute $V$-band magnitude, so if $N^*$ is measured from incomplete data, then the resulting satellite properties will be underestimated. We therefore devise an alternate, simple method of obtaining $N^*$ that is minimally plagued by incompleteness.

We aim to select stars on a clean CMD. Based on the stellar density model, 96\% of stars should be contained within $3r_{\rm ell}$, and while additional completeness could be achieved by extending our selection beyond $3r_{\rm ell}$, we risk introducing foreground contamination. We then select stars that fall in our standard isochrone selection box (0.1\,mag width added in quadrature with photometric uncertainties) that are brighter than $i=22$, where completeness is still high in the \draiii\ region. We count 14 stars that meet these criteria. To fold this into the mass/magnitude estimation method, for each synthetic SSP, we draw a value for $N$ target stars brighter than $i=22$ from a Poisson distribution under the expectation of 14 counts. We estimate that \draiii\ has $M_V = -4.0\pm0.6$ and a stellar mass of roughly 5000\,\Msun.

\draiii\ is the brightest satellite discovered in the UNIONS data analyzed in this work. It sits firmly within the size-luminosity space associated with UFDs so we name it after the Draco constellation. Given the stipulations about shallow photometry discussed in this section, the presence of 21 co-moving stars in {\it Gaia} represents the strongest evidence of \draiii\ being a legitimate, coherent stellar system. Future chemodynamical analysis will be required to confirm \draiii\ as a dark matter-dominated galaxy.

\subsubsection{A Note on the Distance to \draiii}
\label{subsubsec:d3-dist}

Examining the {\it Gaia} CMD in the bottom row of Fig. \ref{fig:cmd-gaia}, we see two members bluer than $(BP-RP)_0 = 0.5$ that may be HB stars. These stars are not reliable for distance estimation with overlaid isochrones due to their proximity to the {\it Gaia} detection limit and subsequent large photometric uncertainties. However, one of these stars is a potential BHB star meeting the criteria in Sec. \ref{subsubsec:bhb} ({\it Gaia} ID ending in 2048). We compute its heliocentric distance to be $90\pm7$\,kpc, which is just barely inconsistent with the isochrone-based distance of $130\pm30$\,kpc. 

We explore whether a closer distance is compatible with the stellar population through visual comparison with isochrones. We fixed the distance to $90$\,kpc and modified the metallicity, finding good agreement with the RGB for a less metal-poor isochrone of [Fe/H] = $-1.5$. If \draiii\ is not as metal-poor as our current model, it may therefore be closer than the current estimate. If this is the case, then we estimate that \draiii\ would be fainter ($M_V \sim -3.4$) and lower stellar mass ($\sim 3000$\,\Msun). However, this BHB-based distance only comes from a single star, and due to the $i$-band incompleteness, we have no reliable photometric information about the SGB or MSTO, two useful features of the stellar population for isochrone-based distance estimates. 
Future spectroscopy (that can anchor the metallicity) and future photometry (that reaches the MSTO) will provide more clarity on the properties of \draiii.

\begin{deluxetable*}{lccccccccccc}
\label{tab:members}
\tabletypesize{\scriptsize}
\tablewidth{0pt}
\tablecaption{Candidate Member Stars for \boovi\ and \draiii, ordered by $r_{\rm ell}$}

\tablehead{
    {\it Gaia} ID & R.A. & Dec. & $g$ & $r$ & $i$ & $G$ & $G_{\rm BP}-G_{\rm RP}$ & $\mu_{\alpha}\cos{\delta}$ & $\mu_{\delta}$ & $r_{\rm ell}$ & $P_{\mathrm{sat}}$ \\
    -- & deg & deg & mag & mag & mag & mag & mag & mas\,year$^{-1}$ & mas\,year$^{-1}$ & 1/\rh & --
}

\startdata
\boovi & & & & & & & & & & & \\
\hline
1377857085156586880 & 235.91521 & 40.00059 & 21.22 & 20.65 & 20.48 & 20.72 & 0.96& $-0.14\pm1.00$ & $1.26\pm1.09$& 0.0 & 0.98 \\
1377857016437104256 & 235.93162 & 39.99558 & -- & 19.93 & 19.69 & 19.99 & 1.17& $0.32\pm0.45$ & $0.04\pm0.47$& 0.7 & 0.96 \\
1377856986373067264 & 235.91273 & 39.97530 & 20.73 & 20.14 & 19.95 & 20.20 & 1.06& $-0.39\pm0.51$ & $0.50\pm0.51$& 1.0 & 0.96 \\
1377856157443591808\tablenotemark{a} & 235.94059 & 39.93902 & 20.74 & 20.81 & 20.88 & 20.78 & 0.38& $-0.32\pm1.10$ & $-0.77\pm1.26$& 2.3 & 0.56 \\
1377858700064322944 & 235.95997 & 40.03104 & 21.09 & 20.69 & 20.53 & 20.71 & 0.50& $-1.40\pm0.86$ & $0.59\pm0.71$& 2.4 & 0.59 \\
1377862140332912512 & 235.86861 & 40.07766 & 21.01 & 20.77 & 20.76 & 20.76 & 0.75& $-0.49\pm1.21$ & $-0.34\pm0.90$& 3.1 & 0.19 \\
1377862140333505152 & 235.86510 & 40.09120 & 21.34 & 20.86 & 20.68 & 20.90 & 1.10& $-0.81\pm1.28$ & $-1.88\pm1.30$& 3.5 & 0.10 \\
\multicolumn{12}{c}{\textemdash\ \textemdash\ \textemdash\ \textemdash\ \textemdash\ \textemdash\ \textemdash\ \textemdash\ \textemdash\ \textemdash\ \textemdash\ \textemdash\ \textemdash\ \textemdash\ \textemdash\ \textemdash\ \textemdash\ \textemdash\ \textemdash\ \textemdash\ \textemdash\ \textemdash\ \textemdash\ \textemdash\ \textemdash\ \textemdash\ \textemdash\ \textemdash\ \textemdash\ \textemdash\ \textemdash\ \textemdash\ \textemdash\ \textemdash\ \textemdash\ \textemdash\ \textemdash\ \textemdash\ \textemdash\ \textemdash\ \textemdash\ \textemdash\ \textemdash\ \textemdash\ \textemdash\ \textemdash} \\
1377849697812833920\tablenotemark{b} & 235.58976 & 39.99980 & 18.76 & 17.94 & 17.68 & 18.02 & 1.28& $-0.50\pm0.11$ & $-0.23\pm0.14$& 13.6 & 0.002 \\
\hline
\hline
\draiii & & & & & & & & & & & \\
\hline
2268532731392831616 & 280.28955 & 75.67528 & 20.62 & 19.88 & 19.63 & 19.94 & 1.21& $0.18\pm0.50$ & $-0.05\pm0.44$& 0.2 & 0.99 \\
2268532804409102720 & 280.26311 & 75.67912 & 20.70 & 19.93 & 19.70 & 20.02 & 1.11& $-0.31\pm0.46$ & $1.15\pm0.41$& 0.2 & 0.95 \\
2268532800112313984 & 280.24107 & 75.68350 & 19.55 & 18.70 & 18.42 & 18.77 & 1.30& $-0.07\pm0.20$ & $0.43\pm0.18$& 0.5 & 0.99 \\
2268532426451439616 & 280.23478 & 75.67185 & 20.51 & -- & 19.43 & 19.76 & 1.13& $0.37\pm0.39$ & $0.14\pm0.34$& 0.7 & 0.98 \\
2268532765752570752 & 280.35176 & 75.67613 & 21.35 & 20.67 & 20.41 & 20.73 & 0.92& $-1.72\pm1.29$ & $0.19\pm1.13$& 1.0 & 0.66 \\
2268533006270754944 & 280.19516 & 75.69706 & 19.86 & 18.99 & 18.73 & 19.09 & 1.27& $-0.21\pm0.25$ & $0.60\pm0.21$& 1.3 & 0.95 \\
2268532323371845248 & 280.27313 & 75.65141 & 21.63 & -- & 20.46 & 20.81 & 1.16& $0.28\pm2.25$ & $1.45\pm2.80$& 1.3 & 0.73 \\
2268532937551282304 & 280.28360 & 75.70506 & 21.47 & 20.28 & 19.88 & 20.38 & 1.32& $1.25\pm0.65$ & $-1.43\pm0.69$& 1.4 & 0.18 \\
2268532319075956480 & 280.31824 & 75.64781 & -- & 19.84 & 19.54 & 19.87 & 1.23& $0.42\pm0.48$ & $0.58\pm0.42$& 1.4 & 0.95 \\
2268521014722050304 & 280.39417 & 75.67758 & 20.42 & 19.46 & 19.16 & 19.56 & 1.35& $0.20\pm0.33$ & $-0.60\pm0.32$& 1.6 & 0.32 \\
2268530979046151296 & 280.18346 & 75.63288 & 20.96 & 20.22 & 19.99 & 20.29 & 1.19& $1.18\pm0.82$ & $-0.16\pm0.59$& 2.9 & 0.47 \\
2268539053584736768 & 280.23699 & 75.74486 & 21.07 & 20.47 & 20.32 & 20.52 & 0.87& $-0.78\pm0.80$ & $0.72\pm0.78$& 3.2 & 0.16 \\
2268533865264234752 & 280.06271 & 75.72990 & 20.58 & 19.94 & 19.69 & 19.97 & 1.32& $-0.37\pm0.43$ & $-0.19\pm0.42$& 3.3 & 0.46 \\
2268531807975603584 & 280.03063 & 75.65969 & 21.35 & -- & 20.40 & 20.73 & 1.17& $-1.55\pm1.04$ & $1.86\pm1.06$& 3.7 & 0.13 \\
2268520258807776896 & 280.55296 & 75.62226 & 21.08 & 20.37 & 20.10 & 20.43 & 1.00& $-0.39\pm0.92$ & $-0.01\pm0.73$& 4.0 & 0.14 \\
2268520812859202048\tablenotemark{b} & 280.59427 & 75.66594 & -- & -- & 20.54 & 20.56 & 0.26& $0.05\pm0.91$ & $-0.59\pm0.71$& 4.2 & 0.63 \\
2268526787158106752 & 280.57505 & 75.69194 & -- & -- & 20.53 & 20.45 & 0.04& $-0.35\pm0.90$ & $1.09\pm0.91$& 4.3 & 0.98 \\
2268531838039626112 & 279.98030 & 75.66125 & 19.35 & -- & 18.32 & 18.62 & 1.26& $0.25\pm0.17$ & $0.28\pm0.16$& 4.4 & 0.44 \\
2268526065603594624 & 280.69058 & 75.68405 & 19.50 & -- & 18.22 & 18.57 & 1.43& $-0.21\pm0.17$ & $0.26\pm0.17$& 5.7 & 0.72 \\
2268532009838326144 & 279.83176 & 75.67467 & -- & -- & 19.08 & 19.41 & 1.23& $0.26\pm0.29$ & $0.55\pm0.28$& 6.2 & 0.28 \\
2268540531053526016 & 280.09703 & 75.82611 & 18.87 & 18.60 & 17.66 & 18.04 & 1.41& $-0.07\pm0.13$ & $0.22\pm0.12$& 7.1 & 0.75 \\
\enddata
Note that all members are given in observed magnitudes, no extinction corrections are made here. Additionally, several member stars do not contain $g$-band data, as explained in App. \ref{app:missing}, and are therefore denoted `--' in the table. Some members of \draiii\ are also missing $r$-band data due to chip gaps and are similarly demarcated.
\tablenotetext{a}{Potential BHB stars used to estimate heliocentric distance}
\tablenotetext{b}{Potential member at 13.6 elliptical \rh\ from the centroid of \boovi}
\end{deluxetable*}

\section{Discussion}
\label{sec:discussion}


\begin{figure}
    \centering
    \includegraphics[width=\linewidth]{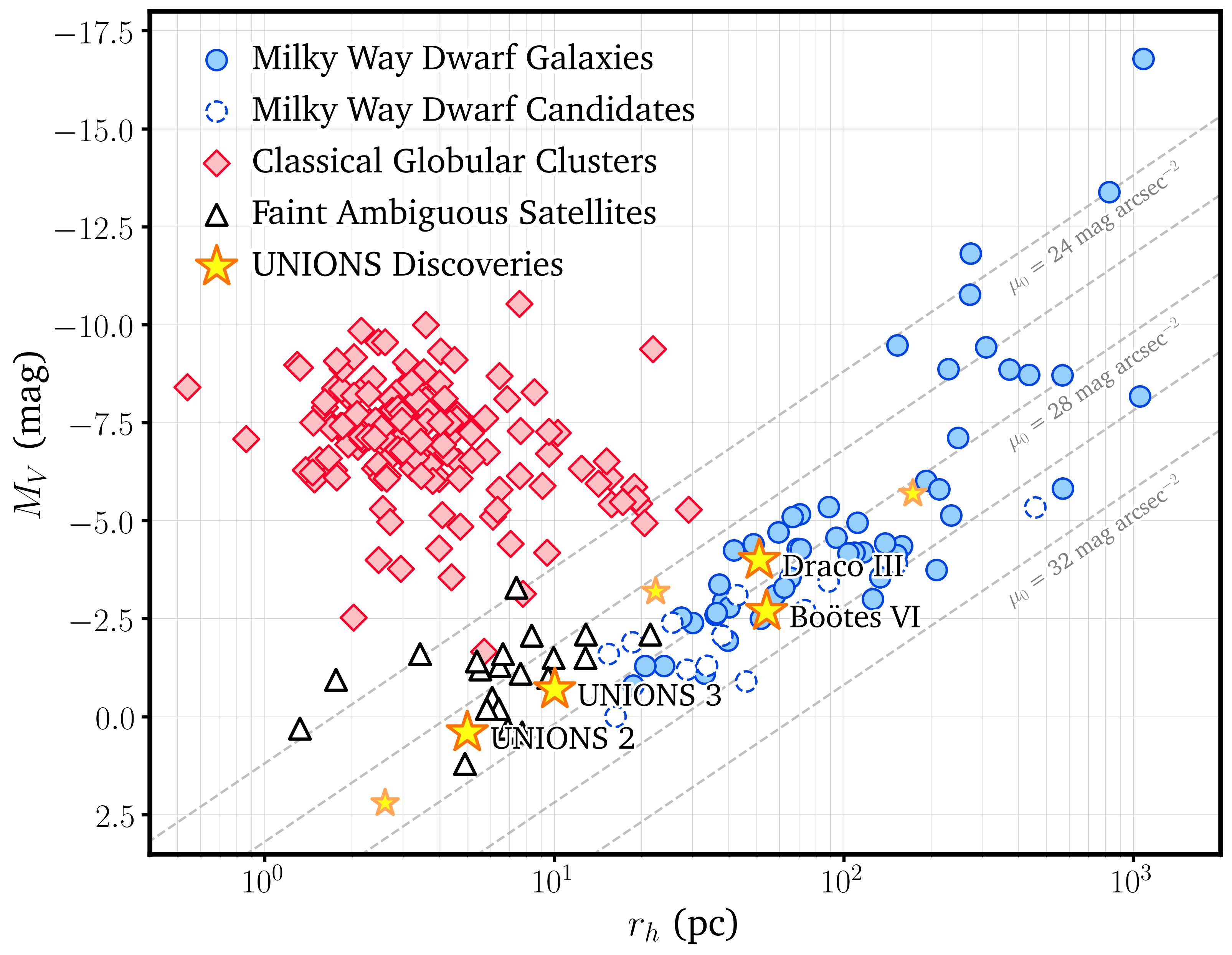}
    \caption{Physical half-light radius (\rh) plotted against total absolute V-band magnitude ($M_V$) for all Milky Way halo satellites. Filled blue circles are all confirmed dwarf galaxies while dashed open blue circles are candidate dwarfs. Red diamonds are all halo globular clusters, and black triangles are all of the faint ambiguous satellites. All data are taken from the Local Volume Database \citep[][as of September 2026]{Pace25}. Large yellow stars are the four new satellites presented in this manuscript, and are labelled as such. The three smaller yellow stars are previous UNIONS discoveries: Pegasus VII, Bo\"otes V, and Ursa Major III/UNIONS 1 in descending order of \rh. Grey dashed lines show constant effective surface brightness assuming a circular light profile.}
    \label{fig:sizelum}
\end{figure}

We have provided initial classifications for each satellite presented in this work based solely on properties derived from deep imaging, namely physical size (\rh) and total absolute $V$-band magnitude ($M_V$) which we present in comparison with all known GCs, Milky Way dwarf galaxy satellites, and  ambiguous UFCSs in Fig. \ref{fig:sizelum}.
\boovi\ and \draiii\ lie comfortably within the distribution of dwarf galaxies at roughly \rh$\ \sim 50$\,pc and $M_V>-5$. \uii\ and \uiii\ are exceptionally faint, fainter than all of the Classical GCs \citep[][2010 edition; $M_V \gtrsim -3$]{Harris96}, but more compact than all confirmed galaxies ($r_\mathrm{h} < 20$\,pc). 

While extendedness is a useful separation criterion for bright systems, the merger of the faint end of the GC luminosity function with the compact end of the dwarf galaxy luminosity function has limited its applicability \citep[e.g.][]{Laevens15-one, Laevens15-three, Conn18}. Additionally, size alone does not provide a physically motivated test for the presence of dark matter. 
Stellar dynamics that imply the presence of a dark matter halo are the gold standard for confirming any satellite as a galaxy in the $\Lambda$CDM context. \citet{Willman12} proposed that a galaxy be defined as a bound stellar system whose behavior cannot be explained by a combination of baryons and Newton's law of gravitation. Under mild assumptions about the anisotropy of stellar orbits, observed line of sight (los) stellar velocities ($v_\mathrm{los}$) can be translated into enclosed dynamical mass \citep{Walker09, Wolf10, Errani18}. \citet{Wolf10} derive $M_{\mathrm{half}}=4G^{-1}\langle\sigma_\mathrm{los}^2\rangle r_{\mathrm{h}}$, where $M_{\mathrm{half}}$ is the mass enclosed in the half-light radius, $\langle\sigma_\mathrm{los}^2\rangle$ is the luminosity-weighted square of the los velocity dispersion, and \rh\ is the 2d projected half-light radius. Every UFD with a well-resolved $\sigma_\mathrm{los}$ is found to have a dynamical mass-to-luminosity ratio ($M_{\mathrm{dyn}}/L$) within 1\rh\ exceeding 50, and sometimes reaching a few 1000 for the most extreme cases \citep[See][and references therein]{Simon19}.

Returning to the mass estimator of \citet{Wolf10}, we can invert the relationship to predict the intrinsic velocity dispersion of each satellite and the velocity precision necessary to probe their dynamics.
For each system we take $M_{\mathrm{half}}$ to be half the total stellar mass, and assume the stellar mass-to-light ratio is 2. We then compute the predicted velocity dispersion for the stellar-only case, a mildly dark matter-dominated case of $M_{\mathrm{dyn}}/L = 50$, and a strongly dark matter-dominated case of $M_{\mathrm{dyn}}/L = 1000$. The results of this exercise are found in Table \ref{tab:dispersion}. Note that we do not attempt to account for the possible presence of unseen baryonic mass in the form of white dwarfs, neutron stars, or black holes in these calculations.

\boovi\ and \draiii\ have 8 and 21 likely member stars, respectively, identified in {\it Gaia} as targets for follow-up spectroscopy. In these candidate galaxies where $M_{\mathrm{dyn}}/L$ is expected to be $\sim 1000$, the predicted $\sigma_{\rm los}$ of 2.7 and 5.1\,\kms\, respectively, should in principle be measurable as individual velocity uncertainties of $<2$\,\kms\ have been readily obtained for stars brighter than the limiting magnitude of {\it Gaia} \citep[e.g.][]{Heiger24, Geha26}.

\uii\ and \uiii, on the other hand, are exemplary of the observational follow-up challenge in the UFCS regime, as neither system contains a candidate member star brighter than $g_0\sim21$, and a spectroscopic sample of $10-15$ stars can only be reached at a depth of $g_0\sim23$. To illustrate the prospects of probing the dark matter content in \uii\ and \uiii\ through stellar spectroscopy, we highlight a recent study of the faintest known satellite, Ursa Major III/UNIONS 1 \citep{Cerny26-uma3u1}, using the DEep Imaging Multi-Object Spectrometer \citep[DEIMOS;][]{Faber03} instrument at the 10-m Keck II telescope, a reliable spectrometer in the northern hemisphere for measuring stellar kinematics in Milky Way satellites \citep[e.g.][]{Martin07, Simon07}. In 4.66 hours of exposure time, \citet{Cerny26-uma3u1} obtained $v_\mathrm{los}$ uncertainties of 2 to 10\,\kms\ for stars in the range $21 < r_0 < 23$. This level of precision may barely be sufficient to resolve $\sigma_\mathrm{los}$ if either system is highly dark matter dominated while mildly dark matter dominated cases would be extremely difficult to robustly establish. 

\begin{deluxetable}{ccccc}
\label{tab:dispersion}
\tabletypesize{\footnotesize}
\tablewidth{0pt}
\tablecaption{Predicted $\sigma_{\rm los}$ based on stellar mass assuming different M$_{dyn}/L$}

\tablehead{
    M$_{dyn}/L$ & \uii & \uiii & \boovi & \draiii \\
    \Msun/\Lsun & km\,s$^{-1}$ & -- & -- & --
}

\startdata
2 & 0.1 & 0.1 & 0.1 & 0.2 \\
50 & 0.6 & 0.6 & 0.6 & 1.1 \\
1000 & 2.6 & 2.7 & 2.7 & 5.1 \\
\enddata
\end{deluxetable}

The challenge is exacerbated by the potential presence of binary stars, which may be dominant in metal-poor environments \citep{Moe19} and whose temporally changing $v_\mathrm{los}$ can artificially inflate the measured velocity dispersion \citep{McConnachie10, Spencer17}. In fact, the most extreme orbital periods can lead to changes on the order of days, necessitating that velocity dispersions are measured from single-epoch spectroscopy, and that multiple epochs are used to weed out binaries.

Taken together, it will require multiple epochs of several-hours-long exposures using stable spectrometers on 10-m telescopes to reliably measure the largest plausible values for the velocity dispersions of \uii\ and \uiii. Concretely demonstrating that either system is devoid of dark matter ($\sigma_{\mathrm{los}} \sim 0.1$\,\kms) is unlikely to be achieved with current instrumentation for measuring radial velocities. 


Projects like the upcoming 39-m Extremely Large Telescope (ELT) at the European Southern Observatory (ESO) may offer a path forward in the coming years. The ArmazoNes high Dispersion Echelle Spectrograph (ANDES) aims to achieve a resolving power of $R\sim 100,000$ and radial velocity stability of better than 100\,m\,s$^{-1}$ \citep{Marconi24, Roederer24}. We use the most updated online version of the exposure time calculator\footnote{\url{https://andes-etc.brera.inaf.it/}, v5.4} to estimate that in a 1 hour observation in ideal conditions, a signal-to-noise ratio ($S/N$) of 7 can be achieved for stars with $V=21$, while it will take 8hrs to achieve $S/N = 5$ at $V=23$. Therefore, high-resolution chemical analyses will be feasible for the brightest stars in both \uii\ and \uiii, and samples of 10-15 stars can in principle be collected, though the time investment will be massive. However, neither \uii\ or \uiii\ will be visible from Cerro Armazones, the southern site of ELT. A northern hemisphere 30-m class telescope will be necessary to carry out these investigations, assuming similar technical specifications for its instruments.

\citet{Willman12} also advocated for the use of stellar metallicities as a secondary avenue for diagnosis, where the dispersion in iron abundance ($\sigma_\mathrm{[Fe/H]}$) is typically observed to be $< 0.1$\,dex in globular clusters but values of $0.2 < \sigma_\mathrm{[Fe/H]} < 0.7$\,dex are measured for dynamically confirmed, low-mass galaxies \citep{Martin07, Simon07, Leaman12, Willman12}. The interpretation is that even the lowest-mass dwarf galaxies must reside in dark matter halos whose gravitational potential wells are deep enough to retain the energetic ejecta from supernovae and AGB stars, facilitating extended star formation and self-enrichment \citep[e.g.][]{Dekel86, Kirby13}.

For some suspected UFDs, stellar velocities have not been obtained with sufficient precision to confidently resolve $\sigma_\mathrm{los}$ and dynamically identify a dark matter halo. However, it has sometimes been possible to measure $\sigma_\mathrm{[Fe/H]} \gtrsim 0.3$\,dex and classify such systems as likely galaxies \citep[e.g.][]{Kirby15, Fritz19, Cerny25}.
For faint ambiguous satellites like \uii\ and \uiii, we find the same observational limitations as dynamical studies where obtaining precise spectroscopic metallicities will be challenging for intrinsically faint stars. However, alternate methods of measuring the metallicity spread may provide a way forward for systems with few bright stars.

Low-resolution spectral observations ($R \sim 1000$) can obtain good $S/N$ for faint stars more quickly than medium or high-resolution instruments and recent studies show promise for [Fe/H] uncertainties on the order of 0.3\,dex \citep{Chiti18, Luna25, Bissonette26}.
For systems with even a single bright star, high-resolution spectroscopy has emerged as a tool for identifying chemical abundance patterns that are unique to dwarf galaxies, such as extremely low barium and strontium abundances \citep[e.g.,][]{Frebel14, Ji19} or the presence of so-called CEMP-no (carbon-rich, extremely metal-poor with no enhancement, over solar ratios, of heavy neutron-capture elements) stars \citep{Norris10, Simon24}.
Narrow band Calcium (Ca) H\&K imaging has been used to measure systemic metallicities and $\sigma_\mathrm{[Fe/H]}$ in Milky Way satellites for large samples of stars derived from photometric metallicities with [Fe/H] uncertainties on the order of $0.2-0.5$\,dex \citep{Longeard20, Fu23, Barbosa25, Jensen26, Taibi26}, though these uncertainties are typically systematic-dominated. 
One potentially important caveat is that carbon-rich stars (like CEMP-no's) have Ca H\&K photometric metallicities that are significantly overestimated \citep{Martin24}, so photometric metallicity dispersions derived from small numbers of stars, as will be the case for UFCSs, should be interpreted with caution.

A viable path forward may be to use a combination of low-resolution spectroscopy, chemical abundance features, and narrow-band imaging to prioritize UFCS objects (such as \uii\ and \uiii) that show possible signs of enrichment for the heavy time investment on medium-to-high resolution spectrographs, and even on future 30-m class telescopes, that will be necessary to determine the first dynamical detection of dark matter for a galaxy with $M_* \lesssim 100$\,\Msun.

\section{Conclusions}
\label{sec:conclusions}

We have presented the results of a matched filter-based search through the internally available {\it gri} UNIONS photometric catalogues and the discovery of four previously unknown Milky Way satellites. Assessing the completeness of Local Group satellites in UNIONS is underway, and we leave the detailed analysis of these results for a future work.

Two systems, \boovi\ \& \draiii, lie firmly in a region of the size-luminosity plane that is unambiguously occupied by UFDs, but both will require targeted spectroscopic follow-up to confirm that they are dark matter dominated. \boovi\ has a particularly low surface brightness and has a potential member star that sits 13.6\rh\ from the system centroid. \draiii\ has 21 members identified through their proper motions that comprise a robust list for a future spectroscopic study, but its age, metallicity, and morphological properties remain poorly constrained due to its identification in a region of UNIONS with poor photometric coverage.

The two remaining systems, \uii\ \& \uiii, are extremely faint ($M_V = +0.4, -0.7$\,mag, respectively) and have no confirmed bright member stars. We have demonstrated that it is extremely challenging to measure their predicted stellar velocity dispersion with current instrumentation if they are mildly dark matter dominated or dark matter free.
We advocate for the use of less time-intensive observational techniques for obtaining first-pass estimates of the metallicity dispersion in extremely faint systems to prioritize the most promising systems that deserve to receive deep spectroscopic follow-ups. 
It is a serious observational challenge to resolve the dark matter-free velocity dispersions for any of the ambiguous UFCSs, a problem that will build in scale as faint ambiguous Milky Way satellites continue to be discovered in current wide-field surveys (e.g. DELVE, UNIONS) and the upcoming programs that will revolutionize our understanding of the Local extragalactic environment such as LSST \citep{Ivezic19}, the \textit{Euclid} Wide survey \citep{Euclid25-overview}, and the Roman High-Latitude Wide-Area Survey \citep{Spergel15}.

\begin{acknowledgments}

We would like to respectfully acknowledge the L\textschwa\textvbaraccent {k}$^{\rm w}$\textschwa\textipa{\ng}\textschwa n Peoples on whose traditional territory the University of Victoria stands and the Songhees, Esquimalt and $\underline{\text{W}}\acute{\text{S}}$ANE$\acute{\text{C}}$ peoples whose relationships with the land continue to this day.

Data collection for this work was conducted at several observing sites atop Maunakea, mentioned individually below. Therefore, the authors wish to recognize and acknowledge the very significant cultural role and reverence that the summit of Maunakea has always had within the Native Hawaiian community. We are most fortunate to have the opportunity to conduct observations from this mountain.

This work is based on data obtained as part of the Canada-France Imaging Survey, a CFHT large program of the National Research Council of Canada and the French Centre National de la Recherche Scientifique. Based on observations obtained with MegaPrime/MegaCam, a joint project of CFHT and CEA Saclay, at the Canada-France-Hawaii Telescope (CFHT) which is operated by the National Research Council (NRC) of Canada, the Institut National des Science de l’Univers (INSU) of the Centre National de la Recherche Scientifique (CNRS) of France, and the University of Hawaii. This research used the facilities of the Canadian Astronomy Data Centre operated by the National Research Council of Canada with the support of the Canadian Space Agency. 
This research is based in part on data collected at Subaru Telescope, which is operated by the National Astronomical Observatory of Japan. 
Pan-STARRS is a project of the Institute for Astronomy of the University of Hawaii, and is supported by the NASA SSO Near Earth Observation Program under grants 80NSSC18K0971, NNX14AM74G, NNX12AR65G, NNX13AQ47G, NNX08AR22G, YORPD20\_2-0014 and by the State of Hawaii. 

This work has made use of data from the European Space Agency (ESA) mission {\it Gaia} (\url{https://www.cosmos.esa.int/gaia}), processed by the {\it Gaia} Data Processing and Analysis Consortium (DPAC, \url{https://www.cosmos.esa.int/web/gaia/dpac/consortium}). Funding for the DPAC has been provided by national institutions, in particular the institutions participating in the {\it Gaia} Multilateral Agreement. 

This work has made use of the Local Volume Database\footnote{\url{https://github.com/apace7/local_volume_database}} \citep{Pace25}.

SS gratefully acknowledges the support of the Natural Sciences and Engineering Research Council of Canada (NSERC).
SS remercie le Conseil de recherches en sciences naturelles et en g\'enie du Canada (CRSNG) de son soutien.

HH is supported by a DFG Heisenberg grant (Hi 1495/5-1), the DFG Collaborative Research Center SFB1491, an ERC Consolidator Grant (No. 770935), and the DLR project 50QE2305.

AV gratefully acknowledges support from the Canadian Institute for Theoretical Astrophysics (CITA) through a CITA National Fellowship and the International Astronomical Union (IAU) and the Gruber Foundation (TGF) through a IAU TGF Fellowship.

SZ acknowledges the support from the Deutsche Forschungsgemeinschaft (DFG) SFB1491. 

\end{acknowledgments}



\facilities{CFHT, PS1, Subaru, {\it Gaia}}

\software{\texttt{astropy} \citep{Astropy13, Astropy18, Astropy22}, \texttt{emcee} \citep{Foreman13}, \texttt{numpy} \citep{Numpy20}, \texttt{scipy} \citep{Scipy20}, \texttt{matplotlib} \citep{Hunter07}}

\appendix
\restartappendixnumbering
\section{Saturation Limits}
\label{app:sat}

\begin{figure}
    \centering
    \includegraphics[width=\linewidth]{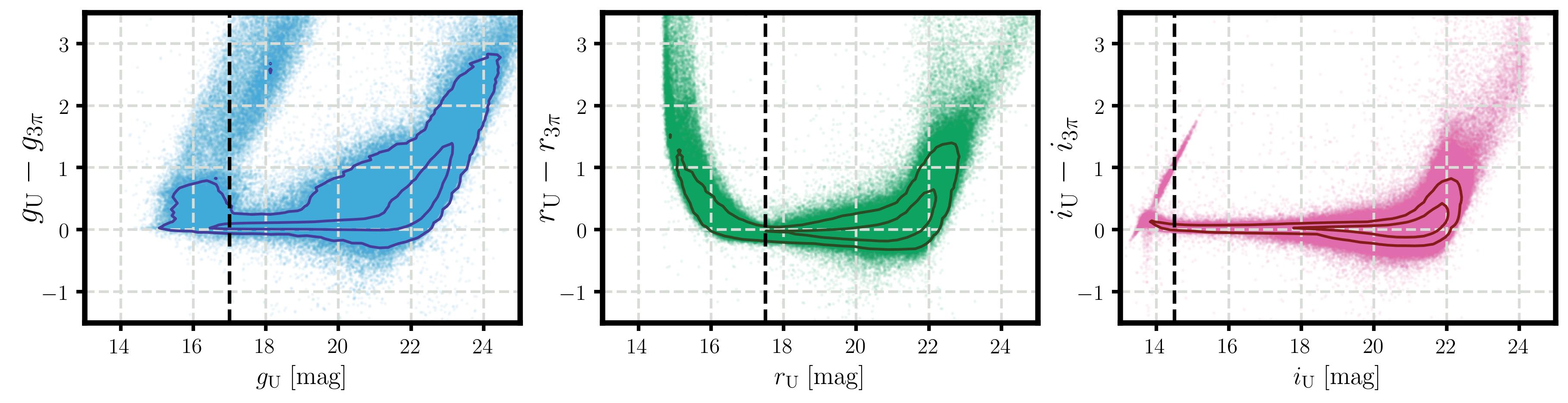}
    \caption{
    Comparison for a set of $>10^6$ common sources between Pan-STARRS 3$\pi$ photometry and the $gri$ bands in UNIONS. We plot the difference between the two bands as a function of the UNIONS photometry and contours enclosing 68 and 95\% of the sources. By eye, we estimate where the UNIONS magnitudes deviate from the Pan-STARRS 3$\pi$ magnitudes, which we take as the saturation limit in each band. This is shown as a vertical dashed black line. Differences in the saturation limit arise between bands because each survey that contributed to UNIONS employs different observing strategies and photometric techniques. {\it Left}: WHIGS $g$-band from Subaru HSC compared with 3$\pi$ where saturation occurs at $g_\mathrm{U} \sim 17$\,mag. {\it Center}: CFIS $r$-band from CFHT MegaCam compared with 3$\pi$ where saturation occurs at $r_\mathrm{U} \sim 17.5$\,mag. {\it Right}: Pan-STARRS $i$-band compared with 3$\pi$ where saturation occurs at $i_\mathrm{U} \sim 14.5$\,mag.
    }
    \label{fig:saturate}
\end{figure}

We investigate the saturation limits in the UNIONS $gri$ photometry by comparing to a large sample of Pan-STARRS $3\pi$ photometry towards the Ursa Minor dwarf spheroidal galaxy. We compare the UNIONS photometry with 2,185,679 $g_{3\pi}$ sources, 1,685,987 $r_{3\pi}$ sources, and 1,689,230 $i_{3\pi}$ sources. We plot the difference between magnitudes from the two surveys as a function of UNIONS photometry in Fig. \ref{fig:saturate} and estimate, by eye, where the UNIONS photometry deviates from $3\pi$ magnitudes at the bright end. This occurs at $g_\mathrm{U} \sim 17$, $r_\mathrm{U} \sim 17.5$, and $i_\mathrm{U} \sim 14.5$. The differences in saturation arise from the UNIONS $i$-band being an extension of the $3\pi$ and employing a similar observing strategy of combing many short exposures. On the other hand, the $g$-band obtained via the WHIGS program at Subaru ($8.2m$ mirror) and the $r$-band obtained via the CFIS program at CFHT ($3.6m$ mirror) both rely on 3-6 exposures of $>100s$ \citep{Gwyn25} and therefore saturate at fainter magnitudes. We apply a boolean cut at these bright limits throughout our analysis to ensure we do not include sources whose magnitudes have been compromised by saturation.

\section{MCMC corner plots}
\label{app:corner}

We present the corner plots resulting from the MCMC-based structural parameter fitting routine presented in Sec. \ref{subsec:structure}. The full chains are also available on \href{https://github.com/simon-et-smith/State-of-the-UNIONS-Candidates/tree/main}{github}.

\begin{figure}
    \centering
    \includegraphics[width=0.9\linewidth]{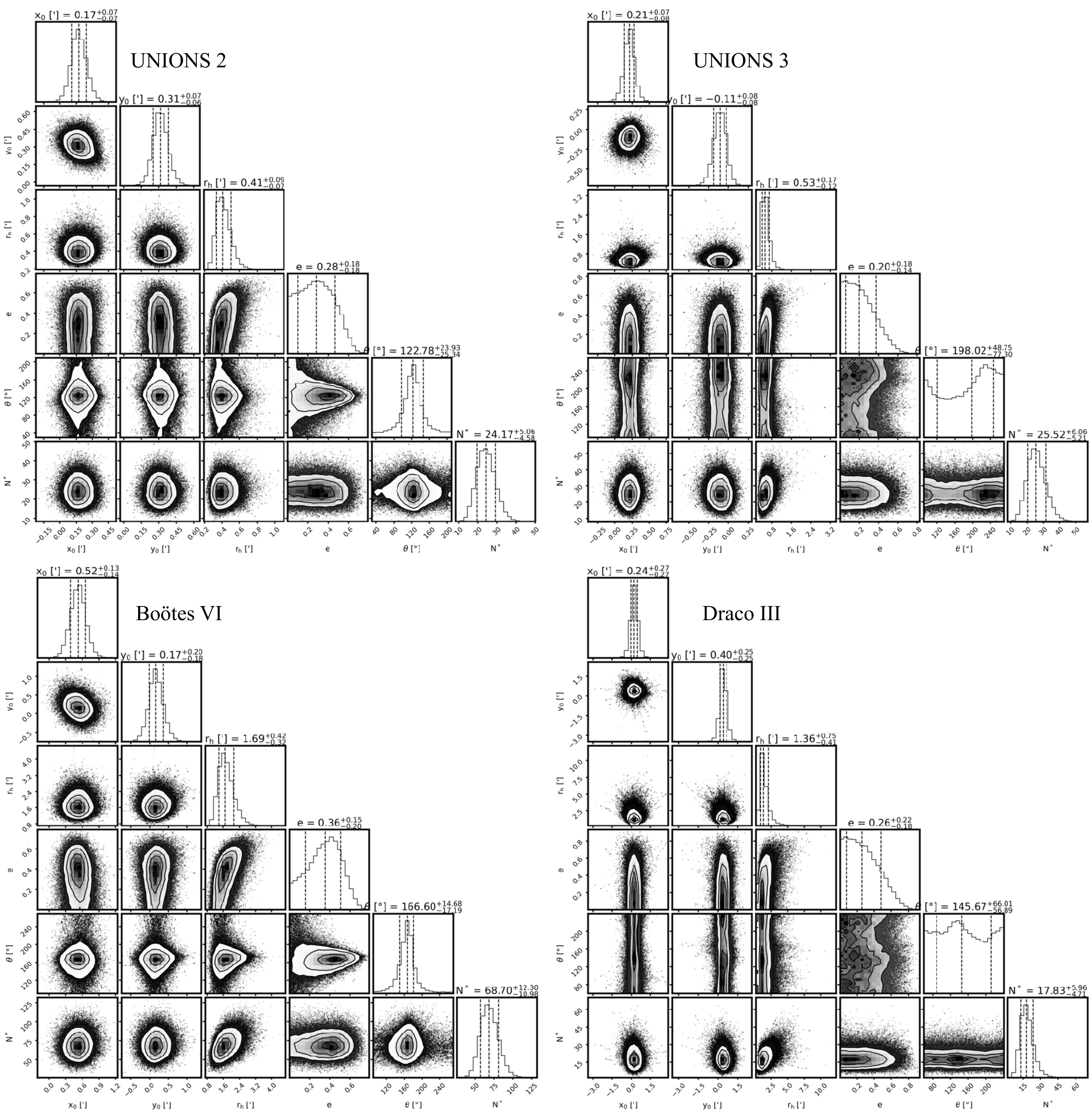}
    \caption{Corner plots for the results of the MCMC-based fitting routine presented in Sec. \ref{subsec:structure}. Each satellite is presented as labeled, with \uii\ in the top left, \uiii\ in the top right, \boovi\ in the bottom left, and \draiii\ in the bottom right. All parameters results are given as the median value with the 16th/84th quantiles of each distribution serving as the lower/upper errorbars, as presented in Table \ref{tab:final-properties}. Note that the ellipticity of both \uiii\ and \draiii\ aren't resolved, which is also noted in Table \ref{tab:final-properties}.}
    \label{fig:mcmc-panel}
\end{figure}

\section{Members missing $g$-band photometry}
\label{app:missing}

\begin{figure*}
    \centering
    \includegraphics[width=0.99\linewidth]{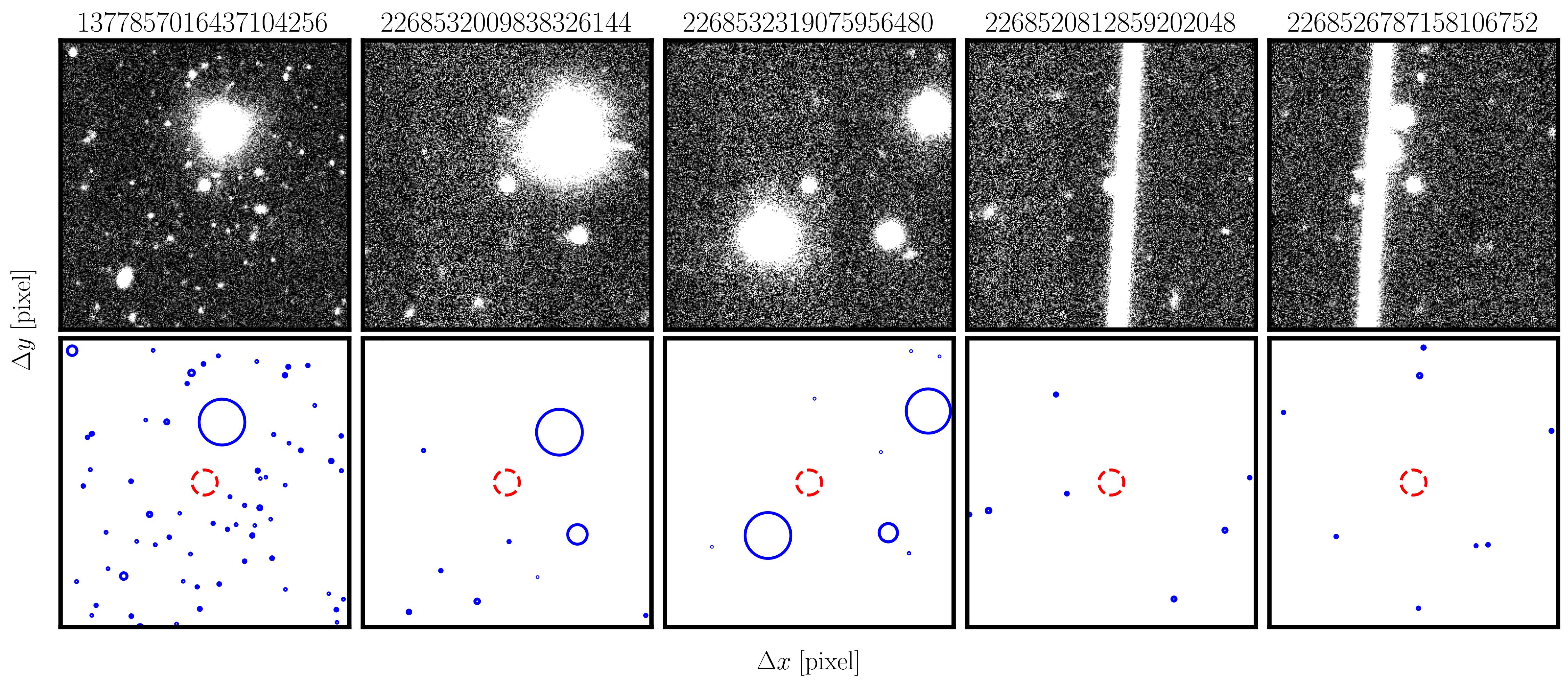}
    \caption{Cutouts (top row) and detected sources (bottom row) for a $300\times300$ pixel region (roughly $1\times1$ arcmin) around the 5 high likelihood members (1 in \boovi, 4 in \draiii) that are missing UNIONS $g$-band photometry. 
    The presence of bright, saturated stars and a man-made satellite track explain why each source was missed by the source detection pipeline.
    In the bottom row, Solid, blue circles are properly detected sources whose size is proportional to their luminosity. The central dashed red circle shows where each member should have been detected.
    The {\it Gaia} source ID is given as the title to each column.}
    \label{fig:memmiss}
\end{figure*}

The stellar membership procedure identified one and four high likelihood member stars in \boovi\ and \draiii, respectively, that lacked $g$-band photometry in the UNIONS catalogue. We plot a $300\times300$ pixel cutout of each source, as well as the locations and relative luminosity of each source detected in that region of sky, in Fig. \ref{fig:memmiss}. In three cases, the presence of bright, saturated stars in the immediate vicinity of the member in question may have caused the source detection pipeline to fail. For two other stars, a bright vertical strip, likely a track from a man-made satellite, similarly seems to have interfered with source detection.

\section{Draco III $i$-band depth}
\label{app:iband}

\begin{figure}
    \centering
    \includegraphics[width=\linewidth]{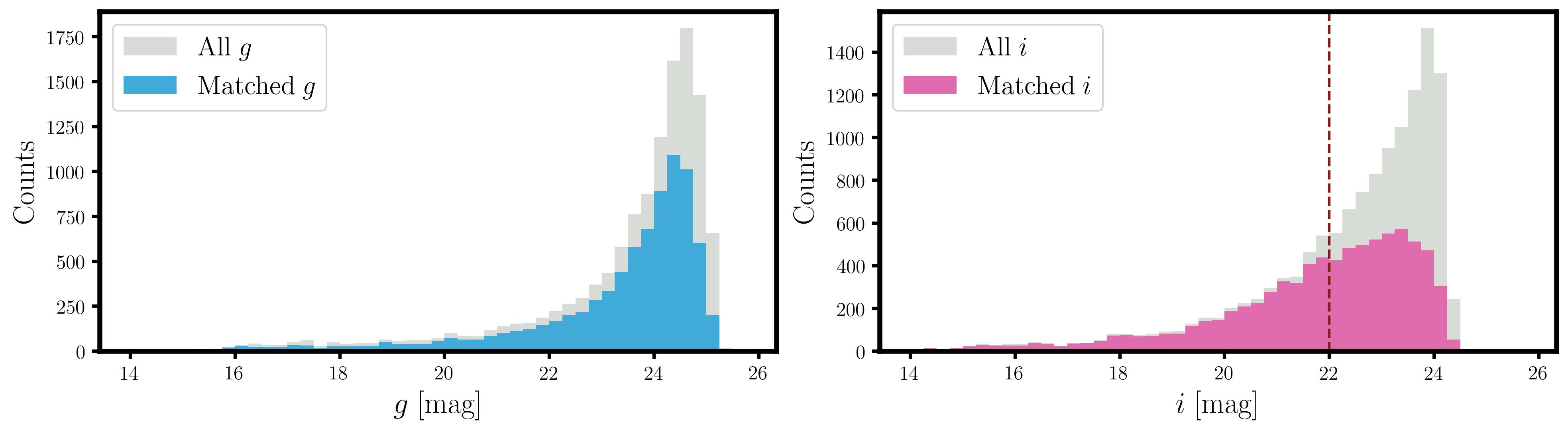}
    \caption{
    Luminosity function for all $gi$ sources in a circle of radius 15 arcminutes about the centroid of \draiii. 
    {\it Left}: $g$-band magnitudes, where the grey histogram shows the distribution of all sources and the blue histogram shows the distribution of sources that are successfully cross-matched to an $i$-band detection in the same region.
    {\it Right}: $i$-band magnitudes, where the grey histogram shows the distribution of all sources and the pink histogram shows the distribution of sources that are successfully cross-matched to a $g$-band detection in the same region. The vertical dark pink line at $i = 22$ shows the magnitude where we estimate the cross-matched distribution to severely deviate from the underlying total distribution.
    }
    \label{fig:dra3-depth}
\end{figure}

The \draiii\ CMD does not appear to be complete down to the local depth of the survey, given assumptions about the nature of the IMF to produce proportionally more stars at progressively lower masses and the presumed distance of \draiii. We examine the properties of the $g$, $i$, and cross-matched $gi$ photometry in a circular region with a radius of 15 arcminutes about the centroid of \draiii\ to investigate the discrepancy. In Fig. \ref{fig:dra3-depth}, we show the magnitude distribution of all sources in grey for both $g$ and $i$. We then show the distribution of $g$ sources that have been successfully cross-matched to an $i$ source in blue and note that it looks like a scaled version of the total distribution. We similarly show the distribution of $i$ sources that have been successfully cross-matched to an $g$ source in pink and find that it has a significantly different shape than the total distribution. We estimate that this deviation becomes severe around $i = 22$, or $i_0 = 21.8$, after being extinction corrected. We therefore use $i = 22$ as our faint limit photometric cut when measuring bulk properties of \draiii\ to mitigate the impact of stellar incompleteness which appears to occur at much shallower magnitudes in this region compared to the regions surrounding other systems in this study.

\bibliography{ref}{}
\bibliographystyle{aasjournal}

\end{document}